\documentclass[journal]{IEEEtran}

\usepackage[utf8]{inputenc}
\usepackage{booktabs}
\usepackage{color}
\usepackage{array}
\usepackage{verbatim}
\usepackage{float}
\usepackage{amsmath}
\usepackage{amsthm}
\usepackage{amssymb}
\usepackage{graphicx}
\usepackage{longtable}
\usepackage[unicode=true,
bookmarks=false,
breaklinks=false,pdfborder={0 0 1},colorlinks=false]
{hyperref}
\hypersetup{
	colorlinks,bookmarksopen,bookmarksnumbered,citecolor=blue,urlcolor=blue}
\usepackage{cite}
\makeatletter
\let\oldforeign@language\foreign@language
\DeclareRobustCommand{\foreign@language}[1]{%
	\lowercase{\oldforeign@language{#1}}}

\let\oldforeign@language\foreign@language
\DeclareRobustCommand{\foreign@language}[1]{%
	\lowercase{\oldforeign@language{#1}}}

\ifCLASSINFOpdf
\else
\fi

\newcommand{\MYfooter}{\smash{
		\hfil\parbox[t][\height][t]{\textwidth}{\centering
			\thepage}\hfil\hbox{}}}

\def\ps@IEEEtitlepagestyle{%
	\def\@oddhead{\parbox[t][\height][t]{\textwidth}{\centering \scriptsize
			Personal use of this material is permitted. Permission from the author(s) and/or copyright holder(s), must be obtained for all other uses. Please contact us and provide details if you believe this document breaches copyrights.\\
			\noindent\makebox[\linewidth]{}
		}\hfil\hbox{}}%
	\def\@evenhead{\scriptsize\thepage \hfil \leftmark\mbox{}}%
	\def\@oddfoot{\parbox[t][\height][l]{\textwidth}{
			\vspace{-20pt}{\rule{\textwidth}{0.4pt}}\\ \footnotesize\underline{To cite this article:}
			{\bf{\footnotesize\textcolor{red}{H. A. Hashim and M. A. Abido, "Location Management in LTE Networks using Multi-Objective Particle Swarm Optimization," Computer Networks, vol. 157, pp. 78-88, 2019.}}} doi: \href{https://doi.org/10.1016/j.comnet.2019.04.009}{10.1016/j.comnet.2019.04.009}\\
			\noindent\makebox[\linewidth]
		}\hfil\hbox{}}%
	\def\@evenfoot{\MYfooter}}

\makeatother
\newtheorem{exam}{Example}

\begin{document}
	\bstctlcite{IEEEexample:BSTcontrol}

	\title{Location Management in LTE Networks using Multi-Objective Particle Swarm Optimization}
	
	\author{Hashim~A.~Hashim$^{*}$, and Mohammad A. Abido\thanks{Manuscript received January 22, 2019; revised April 15, 2019; accepted April 17, 2019.}\thanks{$^{*}$Corresponding author, H. A. Hashim is with the Department of Electrical and Computer Engineering, Western University, London, ON, Canada, N6A-5B9, (e-mail: hmoham33@uwo.ca)}\thanks{M. A. Abido is with Electrical Engineering Department, King Fahd University of Petroleum and Minerals, Dhahran 31261, Saudi Arabia, also M. A. Abido is a Senior Researcher at K.A.CARE Energy Research \& Innovation Center, Dhahran, Saudi Arabia (e-mail: mabido@kfupm.edu.sa)}}
	
	
	\markboth{}{Hashim \MakeLowercase{\textit{et al.}}: Location Management in LTE Networks using MOPSO}
	
	\maketitle
	
	\begin{abstract}
		Long-term evolution (LTE) and LTE-advance (LTE-A) are widely used
		efficient network technologies serving billions of users, since they
		are featured with high spectrum efficiency, less latency, and higher
		bandwidth. Despite remarkable advantages offered by these technologies,
		signaling overhead remains a major issue in accessing the network.
		In particular, the load of signaling is mainly attributed to location
		management. This paper proposes an efficient approach
			for minimizing the total signaling overhead of location management
			in LTE networks using multi-objective particle swarm optimization
			(MOPSO). Tracking area update (TAU) and paging are considered to be
			the main elements of the signaling overhead of optimal location management
			in LTE. In addition, the total inter-list handover contributes significantly
			to the total signaling overhead. However, the total signaling cost
			of TAU and paging is adversely related to the total inter-list handover.
			Hence, two cost functions should be minimized where the first function
			is the total signaling cost of TAU and paging and the second cost
			function is the total signaling overhead. The trade-off between these
			two objectives can be circumvented by MOPSO, which alleviates the
			total signaling overhead. A set of non-dominated solutions on the
		Pareto-optimal front is defined and the best compromise solution is
		presented.
		The  proposed algorithm results in a feasible compromise solution
		between the two objectives, minimizing the signaling overhead, and
		in turn, the consumption of the power battery of the user. The efficacy and the robustness of the proposed algorithm have been
		proven through a large scale environment problem illustrative example.
			The location management in LTE networks using MOPSO best compromise
			solution has been compared to the results obtained by a mixed integer
			non-linear programming (MINLP) algorithm.
	\end{abstract}
	

	\IEEEpeerreviewmaketitle{}

	\section{Introduction}
	
	\IEEEPARstart{O}{ver} the last thirty years cellular networks have been evolving rapidly
	due to the increasing demand for transferring effective and adjustable
	media. In addition, increased capacity is an on-going concern owing
	to continuous growth of the number of users. The rapid development
	in cellular networks, for instance, the long-term evolution (LTE)
	and LTE-advance (LTE-A) allowed to help the user to receive high bandwidth
	and fast connectivity \cite{he2018lte,giluka2018enhanced,xu2018hybrid}.
	Nonetheless, wide dissemination of portable devices and new applications
	lead to higher requirements on networks, since signaling overhead
	directly affects the cellular performance. High levels of signaling
	overhead result in increased computational cost and more power consumption
	\cite{he2018lte}. This could take a form of high load and data traffic
	in the core network which harms the cellular network reliability.
	The development in signaling overhead may tackle the problem of data
	traffic by $50\%$ \cite{ref2}. In addition, LTE and LTE-A have a
	flat IP-based heterogeneous architecture that increases the signaling
	overhead \cite{he2018lte}. Accordingly, the research focus was shifted
	from signaling overhead to handling greater amount of data and increasing
	speed connectivity. In fact, several aspects could lead to signaling
	overhead, however, it can be primarily attributed to user mobility.
	
	Signaling overhead is triggered by transmission or reception of data
	between mobile networks and user equipment (UEs). The performance
	of signaling overhead can be evaluated by two factors, namely, paging
	and tracking area update (TAU). From one side, paging is normally
	actuated by mobility management entity (MME) placing an idle-user
	within the tracking area (TA) of the system. TA is commonly termed
	routing area or location area in other applications. TA is a virtual
	area, which combines a group of cells located within a particular
	area \cite{he2018lte}. The main drawback of TA is the service interruption
	which occurs when a particular cell gets reassigned to a different
	TA. In addition, the edge users located between two or more tracking
	areas (TAs) are subject to ping-pong effect. Tracking area list (TAL)
	is the state-of-the-art extension of TA proposed by the 3rd Generation
	Partnership Project (3GPP). Several TAs can be combined to form TAL.
	Thus, the structure of TA and TAL is nearly similar. However, TAL
	is characterized by greater flexibility and diversity of cells. MME
	defines the location of a user via his/her most recent registered
	location. Accordingly, MME is updated by the last TAL. From the other
	side, TAU is initiated by the user to the MME, this procedure includes
	reporting an update. This update governs transferring the user from
	a cell in one TAL to another TAL. TAL has a prominent role of minimizing
	the total signaling overhead generated by TAU and paging. Thus, the
	design of TAL could allow to adapt various TAs and reduce the ping-pong
	effect. Once the mobile user switches into an idle mode, the signals
	generated by TAU and paging are diminished through TA or, in other
	words, via TAL. Both TAU and paging are employed to track the location
	of the user, and subsequently give a continuous update to the evolved
	packet core (EPC). Consequently, the total signaling cost of TAU and
	paging have a major signaling impact on EPC, allowing them to play
	a vital role for the location management between UEs. An important
	issue to consider is that the inter-list handover resides on three
	factors: UEs, probability of the user to move between cells, and the
	cost of inter-MME. A careful look at the aforementioned discussion
	reveals a bottleneck effect between the total signaling cost of TAU
	and paging from one side, and the total cost of the inter-list handover
	from the other side \cite{razavi2014reducing,aqeeli2017dynamic}.
	Thus, the location management addressed in this work is
	NP-hard, and locating approximate solution
	is NP-hard.
	
	Therefore, this work handles two main factors of optimal location
	management in LTE, the first factor is the total signaling cost of
	TAU and paging, and the second factor is the total inter-list handover.
	The ultimate objective is to minimize the total signaling overhead
	which is a result of paging, TAU and inter-list handover. The problem
	is tackled by optimizing TAL, which is successfully achieved by minimizing
	the total signaling cost of TAU and paging, defined as the first objective,
	and the total inter-list handover - the second objective, consequently
	attenuating the total signaling overhead. It should be remarked that
	TAU and paging are inversely related to the total inter-list handover.
	In an effort to circumvent the above-mentioned challenge, the problem
	can be divided into two-layers. The two-layered approach can be carried
	out by mixed integer non-linear programming (MINLP) \cite{bonami2008algorithmic}
	which divides the problem into two sub-problems. The main disadvantage
	of this approach using MINLP is that one objective is favored over
	the other. By consequence, the final solution obtained by MINLP confines
	the solution on the Pareto-optimal front. Unfortunately, the solution
	obtained will not guarantee the best compromise solution. 
	
	Thus, the main focus is two optimize the two objectives simultaneously.
	According to the conflict between the aforementioned two objectives,
	the trade-off can be achieved through a best compromise among a set
	of non-dominated solutions on the Pareto-optimal front using multi-objective
	particle swarm optimization (MOPSO) \cite{abido2009multiobjective}.
	In this regard, the signaling overhead problem, which is mainly attributed
	to the total signaling of paging, TAU and inter-list handover is studied
	for an arbitrary network. For demonstration and numerical results,
	a large scale network is considered. Numerical results show the robustness
	for the proposed algorithm of finding the best compromise solution
	by solving the trade-off between objective one, which is the total
	signaling cost of TAU and paging, and objective two, which is the
	total inter-list handover. Also, the results guarantee the minimization
	of the signaling overhead problem to lower levels. In addition, the
	proposed algorithm significantly reduces the power consumption of
	the user's device. The contributions of this study can be summarized
	as follows: 
	\begin{enumerate}
		\item The location management problem has been reformulated and approached
		as a multi-objective optimization problem. 
		\item The problem is solved as a minimization problem considering the two
		above-mentioned objectives using MOPSO.
		\item A fuzzy based mechanism to extract the best compromise solution has
		been implemented.
		\item The superiority of the solution obtained by MOPSO over MINLP has been
		proven.
	\end{enumerate}
	The rest of the paper is organized as follows: Related work and an
	overview of mobility management techniques is presented in Section
	\ref{Sec: related}. Section \ref{Sec: sysmod} defines the problem,
	states the system model and formulates the problem. The description
	of MOPSO, the related flow chart, and MOPSO implementation is presented
	in Section \ref{Sec: 4 MO-PSO}. Numerical results and the performance
	of the proposed deployment algorithm compared to MINLP are illustrated
	in Section \ref{Sec: 5 Simulation}. Section \ref{Sec: conclusion}
	includes the summary of the work and the concluding remarks. 
	
	\section{Related Work \label{Sec: related} }
	
	Location management is utilized by various technologies and is popular
	among many researchers \cite{wong2000location,munir2018secure}. Actually,
	location management has motivated scholars to investigate the signaling
	overhead problem. TAU and paging are essential parts of location management
	\cite{wong2000location,ref14}. Number of stereotype techniques aimed
	to attenuate the signaling overhead, for instance (velocity-based,
	timer-based, and movement-based \cite{ref14,ref13,ref15}). The control-plane
	components, for instance, service gateway (SG) or MME, have been employed
	to regulate the signaling load \cite{widjaja2009comparison,kunz2010minimizing,taleb2014supporting}.
	Two schemes of MME pooling, namely the centralized and distributed
	MME pooling schemes have been investigated measuring the load in signaling
	from user mobility \cite{widjaja2009comparison}. However, the study
	in \cite{widjaja2009comparison} did not examine the structure of
	TA and TAL. As introduced in \cite{kunz2010minimizing} a pool of
	SGs assists the TAs in terminating the intermittent connections once
	UE exits one TA and enters another. The minimization of relocation
	frequency has been studied through relocating the data by SGs to support
	high-mobility users \cite{taleb2014supporting}. The shortcoming of
	the approach proposed in \cite{kunz2010minimizing,taleb2014supporting}
	is that it targets only non-idle users, which could weaken the service
	quality through SG re-allocation. 
	
	The signaling overhead problem is mainly attributed to TAL assignment.
	The TAL assignment research can be divided into two categories: static
	and dynamic TAL configuration. TAL design using conventional optimization
	techniques has been introduced through case studies using a linear
	programming (LP) CPLEX (LP-CPLEX) optimizer \cite{ref16,ref17}. The
	results in \cite{ref16,ref17} showed significantly better results
	in terms of signaling overhead compared to the static configuration
	of TAL. The solution proposed in \cite{ref16,ref17} has been extended
	in a new case study with less complex protocol of LP-CPLEX optimizer
	\cite{ref18}. Recently, centralized and distributed MME pooling scheme
	has been employed to attempt addressing the signaling overhead problem
	\cite{aqeeli2017dynamic}. The work in \cite{aqeeli2017dynamic} used
	MINLP optimizer to solve the trade-off between the total signaling
	cost of TAU and paging and the total inter-list handover. It was found
	that the centralized MME pooling scheme outperforms the distributed
	MME pooling scheme in terms of minimum total signaling overhead. It
	is important to recall that 
	the location management in LTE networks is NP-hard and finding appropriate
	solution is NP-hard. Therefore, a critical drawback of the study in
	\cite{aqeeli2017dynamic} is that the algorithm used favored one objective
	over the other one, thus the solution obtained is far from being the
	best compromise solution. 
	
	Particle swarm optimization (PSO) is a global search technique which
	considers the natural behavior of bird swarming \cite{eberhart_new_1995}.
	It has been used effectively to solve: the problem of resource allocation
	and maximization of the system throughput while keeping the minimum
	user rate requirement \cite{du2015using}, sensor deployment in designing
	sensor networks \cite{wu2006energy}, near optimal deployment of sensors
	with Voronoi diagram evaluating the cost function \cite{ab2009wireless},
	and optimal tuning of fuzzy feedback filter \cite{hashim2015_L1_SISO}.
	PSO algorithm introduced in \cite{eberhart_new_1995,abido_optimal_2002}
	is applicable for single objective maximization problems. The multi-objective
	version of PSO is termed MOPSO. Due to the fact that a single optimal
	solution is no longer available in multi-objective context, MOPSO
	is concerned with selecting the best compromise solution from a set
	of non-dominated solutions \cite{abido2009multiobjective,abido2010multiobjective,sun2018attack,hashim2017_L1_MIMO}.
	Moreover, evolutionary techniques have been efficient of solving several
	recent problems, operational aircraft maintenance routing \cite{eltoukhy1,eltoukhy3,eltoukhy5}, directional steering \cite{kamel2017online,kamel2017quad},
	and optimization of membership functions of a fuzzy logic controller
	utilized in control applications \cite{Hash_2015_Comparative,hashim2015_L1_SISO,hashim2017_L1_MIMO,mohamed2014improved}.
	In the next section, the problem of location management in LTE networks
	is formulated to be subsequently solved by MOPSO.
	
	\section{Problem Formulation \label{Sec: sysmod}}
	
\begin{figure}[!h]
		\centering{}\includegraphics[scale=0.22]{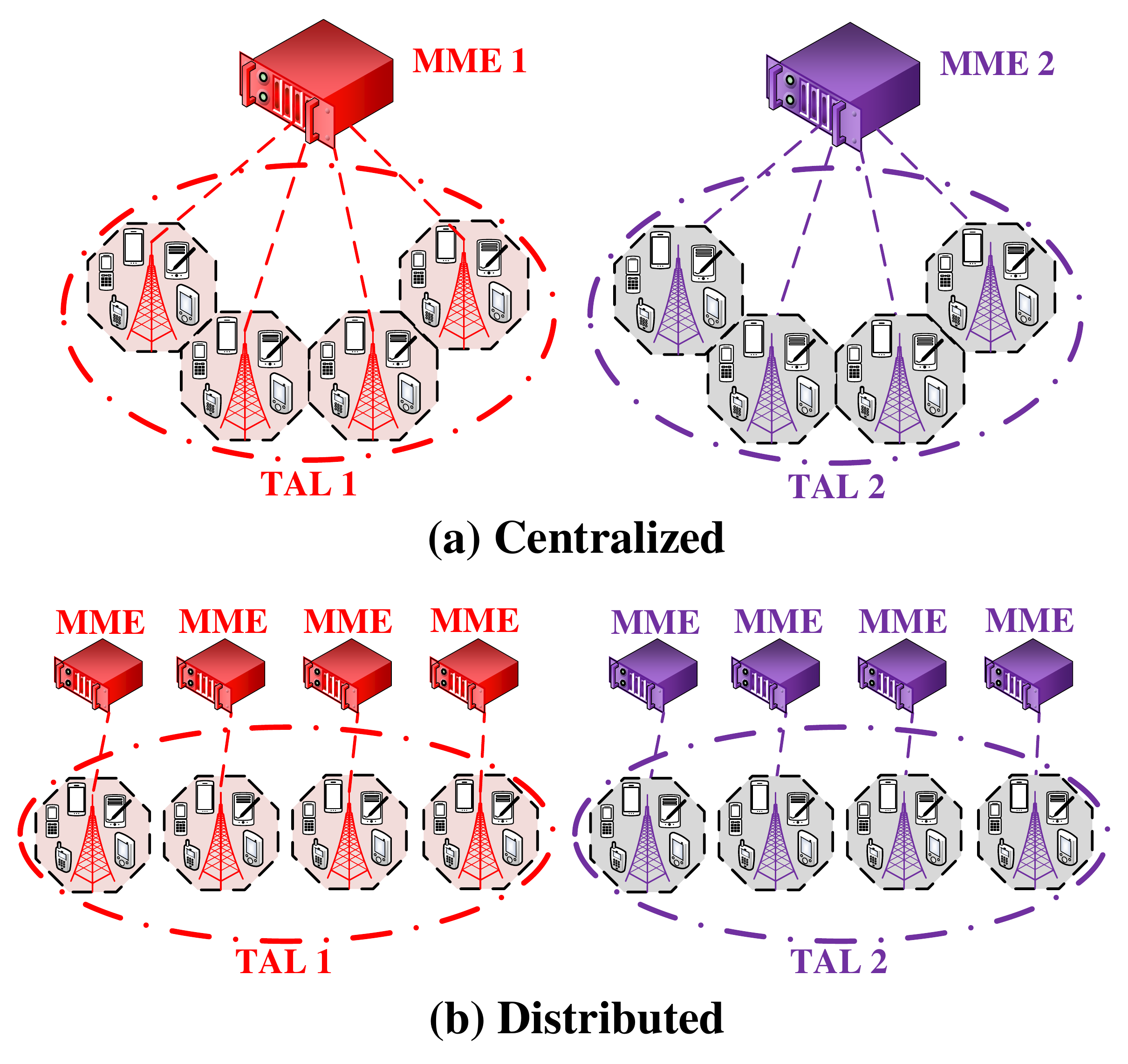}
		\caption{Graphical illustration of the centralized and distributed MME pooling
			scheme: (a) centralized MME pooling scheme; and (b) distributed MME
			pooling scheme}
		\label{fig: fig_centeralized_distributed} 
	\end{figure}

	This study aims to minimize the two main contributing factors to the
	total signaling overhead problem, namely, the total signaling cost
	of TAU and paging and the total inter-list handover. This can be accomplished
	by defining the optimal deployment of overlapping TALs. The total
	signaling overhead problem can be tackled by one of the two popular
	schemes: centralized pooling scheme \cite{razavi2014reducing,widjaja2009comparison}
	or distributed pooling scheme \cite{widjaja2009comparison}. Both
	schemes associate the distribution of TAL with MME. The main difference
	between centralized scheme and distributed one is the method of calculating
	TAU. According to the result of comparison between the two schemes,
	distributed scheme requires extra signal load from MME relocation,
	thus, it can be stated that centralized scheme is more efficient than
	the distributed scheme \cite{razavi2014reducing,aqeeli2017dynamic}.
	Therefore, the problem considered in this section utilizes a centralized
	MME pooling scheme \cite{razavi2014reducing,widjaja2009comparison}.
	Both schemes are illustrated in Figure \ref{fig: fig_centeralized_distributed},
	where the upper portion depicts the centralized MME pooling scheme,
	while the lower portion shows the distributed one.
	
	\subsection{Important Notations}
	
	Table \ref{tab: Table1_MATH_NETWORK} presents math and network notation
	that will be used throughout the paper. Table \ref{tab: Table2_MOPSO_Notations}
	lists notation related to the multi-objective particle swarm optimization
	algorithm.  For any $\boldsymbol{N}\in\mathbb{N}$, $n\in\left[1,\boldsymbol{N}\right]$ indicates that $n$ is an integer number between 1 and $\boldsymbol{N}$. $z\in\left\{ 0,1\right\}$  denotes a binary number or more simple $z\in\mathbb{Z}$. Throughout the paper, $n\in\left[1,\boldsymbol{N}\right]$,
	$k\in\left[1,\boldsymbol{N}\right]$, $\boldsymbol{l}\in\left[1,\boldsymbol{L}\right]$,
	$t\in\left[1,\boldsymbol{\mathcal{T}}\right]$, $j=\left[1,\boldsymbol{\mathcal{P}}\right]$,
    and $i\in\left[1,\boldsymbol{\mathcal{N}}\right]$. Also, $\overline{{\rm rand}}\in\mathbb{R}$ and $\overline{{\rm rand}}\in\left[0,1\right]$.
	
	\begin{table}[!hbp]
		\centering{}\caption{\label{tab: Table1_MATH_NETWORK} Math and Network Notation}
		\begin{tabular}{lcl}
			\toprule 
			$\mathbb{R}^{n\times m}$  & :  & Real $n\times m$ dimensional matrix\tabularnewline
			\midrule 
			$\mathbb{Z}^{n\times m}$  & :  & Binary $n\times m$ dimensional matrix\tabularnewline
			\midrule 
			$\mathbb{N}$  & :  & Set of integer numbers\tabularnewline
			\midrule 
			$^{\top}$ & : & Transpose of a component\tabularnewline
			\midrule 
			$\boldsymbol{N}$  & :  & Number of cells within a list, $\boldsymbol{N}\in\mathbb{N}$\tabularnewline
			\midrule 
			$\boldsymbol{L}$  & :  & Total number of lists, $\boldsymbol{L}\in\mathbb{N}$\tabularnewline
			\midrule 
			$\boldsymbol{l}$  & :  & A single list between $1$ and $\boldsymbol{L}$,
			$\boldsymbol{l}\in\left[1,\boldsymbol{L}\right]$\tabularnewline
			\midrule 
			$\overline{\mathbf{MX}}$  & :  & Maximum number of TAs assigned to list $\boldsymbol{l}$\tabularnewline
			\midrule 
			$\mathbf{H}^{{\rm C}}$  & :  & MME cost relocation over handover process\tabularnewline
			\midrule 
			$\overline{\mathbf{Prob}}_{k,n}$  & :  & Probability of a user to move to cell $n$ from cell $k$\tabularnewline
			\midrule 
			$\mathbf{G}^{{\rm a}}$  & :  & Arrival rate of paging\tabularnewline
			\midrule 
			$\mathbf{G}^{{\rm C}}$  & :  & Paging cost of equipment of a particular user\tabularnewline
			\midrule 
			$\mathbf{G}_{k}^{{\rm T}}$  & :  & Total paging cost of cell $k$\tabularnewline
			\midrule 
			$\mathbf{U}^{{\rm C}}$  & :  & TAU cost of UE moving from one list to another list\tabularnewline
			\midrule 
			$\mathbf{U}_{k}^{{\rm T}}$  & :  & Total signaling cost of TAU of cell $k$\tabularnewline
			\midrule 
			$\overline{\mathbf{UE}}_{k}$  & :  & Total number of UEs served by cell $k$\tabularnewline
			\midrule 
			$\overline{\mathbf{GU}}_{k}^{{\rm T}}$  & :  & Total TAU and paging overhead of cell $k$\tabularnewline
			\midrule 
			$\overline{\mathbf{HC}}_{k}^{\boldsymbol{l}}$  & :  & Inter-list handover rate of user(s) in cell $k$\tabularnewline
			\midrule 
			$\mathbf{O}_{{\rm M}}^{\boldsymbol{l}}$  & $=$  & $\left\{ \begin{array}{rl}
			1, & \text{if list \ensuremath{\boldsymbol{l}} belongs to MME \ensuremath{M},}\\
			0, & \text{otherwise}
			\end{array}\right.$\tabularnewline
			\midrule 
			$\mathbf{Q}_{k}^{{\rm TA}}$  & $=$  & $\left\{ \begin{array}{rl}
			1, & \text{if cell \ensuremath{k} belongs to TA,}\\
			0, & \text{otherwise}
			\end{array}\right.$\tabularnewline
			\midrule 
			$\sigma_{k}^{\boldsymbol{l}}$  & :  & Percentage of usage of list $\boldsymbol{l}$ associated with cell
			$k$\tabularnewline
			\midrule 
			${\rm C}_{k,n}^{\boldsymbol{l}}$  & $=$  & $\left\{ \begin{array}{rl}
			1, & \text{if cells \ensuremath{k} and \ensuremath{n} belong to \ensuremath{\boldsymbol{l}},}\\
			0, & \text{otherwise}
			\end{array}\right.$\tabularnewline
			\bottomrule
		\end{tabular}
	\end{table}
	
	\begin{table}[!hbp]
		\centering{}\caption{\label{tab: Table2_MOPSO_Notations} MOPSO Notation}
		\begin{tabular}{lcl}
			\toprule 
			$x_{i,j}$  & :  & Position of parameter $j$ within particle $i$, $\forall i,j\in\mathbb{N}$\tabularnewline
			\midrule 
			$v_{i,j}$  & :  & Velocity of parameter $j$ within particle $i$, $\forall i,j\in\mathbb{N}$\tabularnewline
			\midrule 
			$\mathcal{X}_{i}$  & :  & Candidate solution of particle $i$, $\forall i,j\in\mathbb{N}$\tabularnewline
			\midrule 
			$\mathcal{V}_{i}$  & :  & Velocity of particle $i$, $\forall i\in\mathbb{N}$\tabularnewline
			\midrule 
			$x_{i,j}^{\star}$ & : & $j$th parameter of $i$th best local particle, $\forall i,j\in\mathbb{N}$\tabularnewline
			\midrule 
			$x_{i,j}^{\star\star}$ & : & $j$th parameter of $i$th best global particle, $\forall i,j\in\mathbb{N}$\tabularnewline
			\midrule 
			$\boldsymbol{\mathcal{P}}$  & : & Number of parameters within 1 particle, $\boldsymbol{\mathcal{P}}\in\mathbb{N}$ \tabularnewline
			\midrule 
			$\boldsymbol{\mathcal{N}}$  & :  & Number of particles within 1 population, $\boldsymbol{\mathcal{N}}\in\mathbb{N}$ \tabularnewline
			\midrule 
			$\boldsymbol{\mathcal{T}}$ & : & Total number of iterations/generations, $\boldsymbol{\mathcal{T}}\in\mathbb{N}$ \tabularnewline
			\midrule 
			$t$ & : & Iteration/generation number, $t\in\mathbb{N}$ , $t\in\left[1,\boldsymbol{\mathcal{T}}\right]$\tabularnewline
			\midrule 
			$\alpha\left(t\right)$ & : & Inertia factor at iteration/generation $t$, $\alpha\left(t\right)\in\mathbb{R}$\tabularnewline
			\midrule 
			$\mathcal{S}_{i}^{\star}$ & :  & Non-dominated local set of $i$th particle, $\forall i\in\mathbb{N}$\tabularnewline
			\midrule 
			$\mathcal{S}^{\star\star}$ & :  & Non-dominated global set\tabularnewline
			\midrule 
			$\mathbf{S}^{{\rm L}}$ & :  & Size of non-dominated local set\tabularnewline
			\midrule 
			$\mathbf{S}^{{\rm G}}$ & :  & Size of non-dominated global set\tabularnewline
			\midrule 
			$\boldsymbol{\mathcal{J}}^{1}$  & :  & Objective function 1 to be minimized, $\boldsymbol{\mathcal{J}}^{1}\in\mathbb{R}$ \tabularnewline
			\midrule 
			$\boldsymbol{\mathcal{J}}^{2}$  & :  & Objective function 2 to be minimized, $\boldsymbol{\mathcal{J}}^{2}\in\mathbb{R}$ \tabularnewline
			\midrule 
			$\overline{{\rm rand}}$ & :  & A real random number between 0 and 1\tabularnewline
			\bottomrule
		\end{tabular}
	\end{table}

	\subsection{Preliminaries and Design Hypothesis}
	
	The cells are allocated within every TAL through the centralized MME
	pooling scheme. Every TA refers to a single cell, hence, TA and cell
	will be used interchangeably. The centralized MME pooling scheme is
	depicted in the upper portion of Figure \ref{fig: fig_centeralized_distributed}.
	The model includes two layers of assignments. The first layer considers
	the Cells-to-TALs/MMEs assignment associated with the core network.
	The second layer presents the TALs/MMEs to UEs assignment, which relates
	to every cell within the system. The first layer of the system evaluates
	the UEs mobility pattern among the cells, then it allocates the cells
	in TALs/MMEs with the goal of attenuating the total signaling overhead
	caused by TAU and paging. In the second layer, a number of TALs is
	positioned at a portion of UEs termed $\sigma_{k}^{\boldsymbol{l}}$.
	The selection of $\sigma_{k}^{\boldsymbol{l}}$ defines the usage
	portion of every single TAL/MME per cell which leads to interference
	of TALs through the cells. This technique guarantees a broader assortment
	of cells in the list.
	
	\subsubsection{Cell/TA-to-TAL/MME Assignment}
	
	The allocation of cell/TA-to-TAL/MME in the centralized pooling scheme
	is the main objective of the presented model. Every TAL refers to
	a certain one-to-one MME basis. The allocation of Cells-to-TAL/MME
	is given through a binary decision variable ${\rm C}_{k,n}^{\boldsymbol{l}}\in\mathbb{Z}$
	for all $\boldsymbol{l},k,n\in\mathbb{N}$, $k,n\in\left[1,\boldsymbol{N}\right]$
	and $\boldsymbol{l}\in\left[1,\boldsymbol{L}\right]$. In order to
	proceed with the model and problem formulation, it is necessary to
	present the following example.
	\begin{exam}
		\label{exa:Examp1}As an illustrative example of ${\rm C}_{k,n}^{\boldsymbol{l}}\in\mathbb{Z}$
		allocation in the search space, define ${\rm C}_{k,n}^{\boldsymbol{l}}$
		to be a cell-to-TAL assignment. For one list ($\boldsymbol{l}=1$),
		define the maximum number of cells/TAs to be $\bar{\mathbf{n}}=3$
		with $\boldsymbol{N}=4$ MMEs or lists. Accordingly, the total number
		of possible combinations in matrix $\bar{\boldsymbol{\mathcal{L}}}^{1}={\rm C}^{\boldsymbol{l}}$
		is $3\times4$, where ${\rm C}_{k,n}^{\boldsymbol{l}}$ is a binary
		decision created in the system. One can specify the decision variables
		in list 1 as
		\begin{align}
		\bar{\boldsymbol{\mathcal{L}}}^{1} & =\begin{bmatrix}1 & 0 & 1 & 1\\
		0 & 0 & 0 & 0\\
		1 & 0 & 1 & 1\\
		1 & 0 & 1 & 1
		\end{bmatrix}=\begin{bmatrix}{\rm C}_{1,1}^{1} & {\rm C}_{1,2}^{1} & {\rm C}_{1,3}^{1} & {\rm C}_{1,4}^{1}\\
		{\rm C}_{2,1}^{1} & {\rm C}_{2,2}^{1} & {\rm C}_{2,3}^{1} & {\rm C}_{2,4}^{1}\\
		{\rm C}_{3,1}^{1} & {\rm C}_{3,2}^{1} & {\rm C}_{3,3}^{1} & {\rm C}_{3,4}^{1}\\
		{\rm C}_{4,1}^{1} & {\rm C}_{4,2}^{1} & {\rm C}_{4,3}^{1} & {\rm C}_{4,4}^{1}
		\end{bmatrix}\label{eq:example}
		\end{align}
	\end{exam}
	Equation \eqref{eq:example} in Example \ref{exa:Examp1} shows that
	$\bar{\boldsymbol{\mathcal{L}}}^{1}$is symmetric with $\bar{\boldsymbol{\mathcal{L}}}^{1}=\left(\bar{\boldsymbol{\mathcal{L}}}^{1}\right)^{\top}\in\mathbb{Z}^{4\times4}$.
	Also, it demonstrates that number of nonzero rows/columns has to be
	3. In addition, number of ones in any given nonzero row/column has
	to be 3 with one zero row/column. Therefore, a general cell-to-TAL
	assignment with $\boldsymbol{L}$ lists and $\boldsymbol{N}$ MMEs
	or lists can be expressed by
	
	\begin{align}
	\bar{\boldsymbol{\mathcal{L}}}^{1} & =\left[\begin{array}{c}
	\mathcal{L}_{1}^{1}\\
	\mathcal{L}_{2}^{1}\\
	\vdots\\
	\mathcal{L}_{\boldsymbol{N}}^{1}
	\end{array}\right]=\begin{bmatrix}{\rm C}_{1,1}^{1} & {\rm C}_{1,2}^{1} & \dots & {\rm C}_{1,\boldsymbol{N}}^{1}\\
	{\rm C}_{2,1}^{1} & {\rm C}_{2,2}^{1} & \dots & {\rm C}_{2,\boldsymbol{N}}^{1}\\
	\vdots & \vdots & \ddots & \vdots\\
	{\rm C}_{\boldsymbol{N},1}^{1} & {\rm C}_{\boldsymbol{N},2}^{1} & \dots & {\rm C}_{\boldsymbol{N},\boldsymbol{N}}^{1}
	\end{bmatrix}\nonumber \\
	\vdots\hspace{0.5em} & =\hspace{1em}\hspace{1em}\vdots\hspace{1em}\hspace{1em}\hspace{1em}\hspace{1em}\hspace{1em}\hspace{1em}\hspace{1em}\hspace{1em}\vdots\nonumber \\
	\bar{\boldsymbol{\mathcal{L}}}^{\boldsymbol{L}} & =\left[\begin{array}{c}
	\mathcal{L}_{1}^{\boldsymbol{L}}\\
	\mathcal{L}_{2}^{\boldsymbol{L}}\\
	\vdots\\
	\mathcal{L}_{\boldsymbol{N}}^{\boldsymbol{L}}
	\end{array}\right]=\begin{bmatrix}{\rm C}_{1,1}^{\boldsymbol{L}} & {\rm C}_{1,2}^{\boldsymbol{L}} & \dots & {\rm C}_{1,\boldsymbol{N}}^{\boldsymbol{L}}\\
	{\rm C}_{2,1}^{\boldsymbol{L}} & {\rm C}_{2,2}^{\boldsymbol{L}} & \dots & {\rm C}_{2,\boldsymbol{N}}^{\boldsymbol{L}}\\
	\vdots & \vdots & \ddots & \vdots\\
	{\rm C}_{\boldsymbol{N},1}^{\boldsymbol{L}} & {\rm C}_{\boldsymbol{N},2}^{\boldsymbol{L}} & \dots & {\rm C}_{\boldsymbol{N},\boldsymbol{N}}^{\boldsymbol{L}}
	\end{bmatrix}\label{eq: Stage1_eq1}
	\end{align}
	The assignment of the maximum number of cells within a list has to
	satisfy the following expression
	\begin{equation}
	\sum\limits _{k=1}^{\boldsymbol{N}}\sum\limits _{n,n\neq k}^{\boldsymbol{N}}{\rm C}_{k,n}^{\boldsymbol{l}}\leqslant\overline{\mathbf{MX}},\hspace{1em}\forall\boldsymbol{l}\in\left[1,\boldsymbol{L}\right]\label{eq: Stage1_eq2}
	\end{equation}
	where $\overline{\mathbf{MX}}$ denotes maximum number of TAs in a
	list. Also, according to the above-mentioned discussion and Example
	\ref{exa:Examp1}, the cell/TA assignment in every TAL takes the following
	form
	
	\begin{equation}
	{\rm C}_{k,n}^{\boldsymbol{l}}={\rm C}_{n,k}^{\boldsymbol{l}},\hspace{1em}\forall\boldsymbol{l}\in[1,\boldsymbol{L}],\text{ and }k,n,k\neq n\in\left[1,\boldsymbol{N}\right]\label{eq: Stage1_eq3}
	\end{equation}

	\subsubsection{TAL-to-UE}
	
	According to the fact that any TAL associated with MME could be assigned
	to more than one cell, the assigned cell has not to violate the MME
	load of any nonzero $n$ column, which could be achieved by
	
	\begin{equation}
	\sum\limits _{k=1}^{\boldsymbol{N}}\sigma_{k}^{\boldsymbol{l}}\cdot{\rm C}_{k,n}^{\boldsymbol{l}}=1,\hspace{1em}\forall\:\boldsymbol{l}\in\left[1,\boldsymbol{L}\right]\label{eq: Stage2_eq1}
	\end{equation}
	where $\sigma_{k}^{\boldsymbol{l}}$ denotes the percentage of usage
	of list $\boldsymbol{l}$ associated with cell $k$.
	
	\subsection{Total Signaling of TAU and paging }
	
	The signaling cost of TAU and paging is a fundamental consideration
	when placing an idle-user in a cellular network \cite{ref14,ref15}.
	In fact, the total signaling overhead created by TAU is a result of
	a user movement from a cell in one TAL to a different TAL. Whereas,
	the total signaling overhead created by paging is due to a necessity
	to locate a user via the core network. Since the total signaling overhead
	cost is inversely related to the size of TAL, there is negative correlation
	between the total signaling created by TAU and paging, and the size
	of TAL. This is the case due to the fact that the greater the number
	of cells within a certain list the lower is the probability for a
	user to move between lists. 
	
	\subsection{Problem Description and Cost Functions}
	
	This subsection presents the centralized MME pooling problem \cite{razavi2014reducing,widjaja2009comparison,aqeeli2017dynamic}.
	The objective is to minimize the signaling overhead, which can be
	accomplished by minimizing two objective functions, namely, the total
	signaling cost of TAU and paging, and the total inter-list handover.
	The total signaling cost of TAU of cell $k$ for any $n$ nonzero column is equivalent to{\small{}
		\begin{equation}
		\begin{split}\mathbf{U}_{k}^{{\rm T}}= & \overline{\mathbf{UE}}_{k}\cdot\overline{\mathbf{Prob}}_{k,n}\cdot\mathbf{U}^{{\rm C}}\cdot\left[\sum\limits _{\boldsymbol{l}=1}^{\boldsymbol{L}}\mathbf{H}^{{\rm C}}\cdot\mathbf{O}_{{\rm M}}^{\boldsymbol{l}}\cdot\sigma_{k}^{\boldsymbol{l}}\left(1-{\rm C}_{k,n}^{\boldsymbol{l}}\right)\right],\\
		& \hspace{1em}\forall\:k,n,k\neq n\in\left[1,\boldsymbol{N}\right]
		\end{split}
		\label{centralized1}
		\end{equation}
	}with $\overline{\mathbf{UE}}_{k}$ being the total number of UEs
	served by cell $k$, $\mathbf{U}^{{\rm C}}$ being the TAU cost of
	UE moving from one list to another, $\mathbf{H}^{{\rm C}}$ being
	the MME cost relocation over handover process, and $\mathbf{O}_{{\rm M}}^{\boldsymbol{l}}$
	defining whether list $\ensuremath{\boldsymbol{l}}$ belongs to MME
	$\ensuremath{M}$ or not. The total cost of paging of cell $k$ for any $n$ nonzero column can
	be evaluated by
	\begin{equation}
	\begin{split}\mathbf{G}_{k}^{{\rm T}}= & \mathbf{G}^{{\rm a}}\cdot\mathbf{G}^{{\rm C}}\cdot\left[\sum\limits _{\boldsymbol{l}=1}^{\boldsymbol{L}}\overline{\mathbf{UE}}_{k}\cdot\sigma_{k}^{\boldsymbol{l}}+\sum\limits _{\boldsymbol{l}=1}^{\boldsymbol{L}}\sum\limits _{n,k\neq n}^{\boldsymbol{N}}\overline{\mathbf{UE}}_{n}\cdot{\rm C}_{k,n}^{\boldsymbol{l}}\cdot\sigma_{n}^{\boldsymbol{l}}\right]\\
	& \hspace{4em}\hspace{1em}\forall\:k\in\left[1,\boldsymbol{N}\right]
	\end{split}
	\label{paging1}
	\end{equation}
	where $\mathbf{G}^{{\rm a}}$ denotes the arrival rate of paging and
	$\mathbf{G}^{{\rm C}}$ refers to the paging cost of equipment of
	a particular user. Recall the constraints in Equations \eqref{eq: Stage1_eq1},
	\eqref{eq: Stage1_eq2}, \eqref{eq: Stage1_eq3}, and \eqref{eq: Stage2_eq1}
	\begin{equation}
	\sum\limits _{k=1}^{\boldsymbol{N}}\sigma_{k}^{\boldsymbol{l}}\cdot{\rm C}_{k,n}^{\boldsymbol{l}}=1,\hspace{1em}\forall\:\boldsymbol{l}\in\left[1,\boldsymbol{L}\right]\label{sigma1}
	\end{equation}
	\begin{equation}
	\sum\limits _{k,n,k\neq n}^{\boldsymbol{N}}{\rm C}_{k,n}^{\boldsymbol{l}}\leqslant\overline{\mathbf{MX}},\hspace{1em}\forall\:\boldsymbol{l}\in\left[1,\boldsymbol{L}\right]\label{kappa1}
	\end{equation}
	\begin{equation}
	{\rm C}_{k,n}^{\boldsymbol{l}}={\rm C}_{n,k}^{\boldsymbol{l}},\hspace{1em}\forall\boldsymbol{l}\in[1,\boldsymbol{L}],\:k,n,k\neq n\in\left[1,\boldsymbol{N}\right]\label{equi1}
	\end{equation}
	\begin{equation}
	0\leq\sigma_{k}^{\boldsymbol{l}}\leq1\label{a1}
	\end{equation}
	\begin{equation}
	\mathbf{O}_{{\rm M}}^{\boldsymbol{l}}\in\left\{ 0,1\right\} \label{b1}
	\end{equation}
	\begin{equation}
	{\rm C}_{k,n}^{\boldsymbol{l}}\in\left\{ 0,1\right\} \label{c1}
	\end{equation}
	where Equations \eqref{a1}, \eqref{b1}, and \eqref{c1} refer to
	boundary constraints. Accordingly, the total signaling cost of TAU
	and paging of the $i$th element is defined by
	\begin{equation}
	\overline{\mathbf{GU}}_{k}^{{\rm T}}=\mathbf{G}_{k}^{{\rm T}}+\mathbf{U}_{k}^{{\rm T}},\hspace{1em}\forall\:k\in\left[1,\boldsymbol{N}\right],k\in\mathbb{N}\label{eq:sub-problem2}
	\end{equation}
	and the first objective function is given by
	\begin{equation}
	\boldsymbol{\mathcal{J}}^{1}=\sum\limits _{k=1}^{\boldsymbol{N}}\overline{\mathbf{GU}}_{k}^{{\rm T}}\label{eq:sub-problem2-main_obj_2}
	\end{equation}
	Hence, the total signaling overhead introduced by TAU and paging could
	be minimized through the minimization of $\boldsymbol{\mathcal{J}}^{1}$
	in Equation \eqref{eq:sub-problem2-main_obj_2}. On the other side,
	the total inter-list handover of the $k$th element for any $n$ nonzero column is
	\begin{equation}
	\overline{\mathbf{HC}}_{k}^{\boldsymbol{l}}=\sum\limits _{\boldsymbol{l}=1}^{\boldsymbol{L}}\overline{\mathbf{UE}}_{k}\cdot\overline{\mathbf{Prob}}_{k,n}\cdot\left(1-{\rm C}_{k,n}^{\boldsymbol{l}}\right)\label{eq:sub-problem1}
	\end{equation}
	with the second objective function being defined by
	\begin{equation}
	\boldsymbol{\mathcal{J}}^{2}=\sum\limits _{k=1}^{\boldsymbol{N}}\overline{\mathbf{HC}}_{k}^{\boldsymbol{l}}\label{eq:sub-problem1-main_obj_1}
	\end{equation}
	such that the minimization of total signaling overhead could be achieved
	by diminishing the total inter-MME reallocation cost, which is defined
	by $\boldsymbol{\mathcal{J}}^{2}$ in Equation \eqref{eq:sub-problem1-main_obj_1}.
	The fluid flow model is popular and frequently employed to resemble
	the mobility behavior of a user in a system \cite{razavi2014reducing}.
	The fluid flow model presents the UEs traffic outflow rate of an enclosed
	area. In this specific case enclosed area refers to a single cell.
	For a given cell $k$, let $\overline{PM}$ denote its perimeter,
	$\rho_{k}$ refer to UE density of the cell, and $v$ be the UE average
	velocity. Hence, the average of cell crossings with respect to time
	is defined by
	\begin{equation}
	\dfrac{\rho_{k}\cdot\overline{PM}\cdot v}{\pi}\label{fluid flow}
	\end{equation}
	implying that the cells are hexagonal-shaped with length $L^{{\rm H}}$,
	which means that $\overline{PM}=6L^{{\rm H}}$. 
	
	According to the above-mentioned discussion and the problem statement,
	the two objectives $\boldsymbol{\mathcal{J}}^{1}$ and $\boldsymbol{\mathcal{J}}^{2}$
	are inversely related. A weighting scheme can be applied to the two
	conflicting objectives \cite{hashim2015_L1_SISO}. However, there
	is no guarantee that the obtained solution would be the best compromise
	solution \cite{hashim2017_L1_MIMO}.
	
	\subsection{Overview of model decomposition using MINLP}
	
	The problem formulation of the model mentioned above is NP-hard, thus,
	the problem can be divided into two sub-problems. One sub-problem
	may consider the cell-to-TAL/MME assignment which is allocated periodically
	to minimize the inter-list handover rate of UEs from one cell to another.
	This sub-problem mainly includes Equations \eqref{kappa1}, \eqref{equi1},
	\eqref{eq:sub-problem1} and \eqref{eq:sub-problem1-main_obj_1}.
	The second sub-problem could assume the set of TALs/MMEs given and,
	thereby, using the given data one can find the optimum usage ratio
	for each TAL/MME. The other sub-problem mainly includes Equations
	\eqref{a1}, \eqref{c1}, \eqref{centralized1}, \eqref{paging1},
	\eqref{eq:sub-problem2} and \eqref{eq:sub-problem2-main_obj_2}.
	For a full description of the previously mentioned two sub-problems
	using MINLP visit \cite{bonami2008algorithmic,aqeeli2017dynamic}.
	A scrutiny look at the final solution obtained by MINLP, one can find
	that the solution is confined on the Pareto-optimal front. However,
	one objective is strictly favored the optimum solution over the other
	one, which is a significant weakness. Thus, the best compromise solution
	cannot be achieved using this method. Consequently, a random artificial
	search technique such as optimization algorithms based on the social
	behavior of animals could help in approaching a set of solutions on
	the Pareto-optimal of the NP-hard problem \cite{Hashim_2016_Network}.
	In that case, the two above-mentioned sub-problems would be solved
	simultaneously through the optimization algorithm and the best compromise
	solution would be defined. 
	
	\section{Multi-objective Particle Swarm Optimization\label{Sec: 4 MO-PSO}}
	
	\subsection{Particle Swarm Optimization}
	
	Particle Swarm Optimization (PSO) is considered to be an evolutionary
	heuristic technique. This technique imitates the cognitive and social
	behavior of animals in their natural environment such as fish schooling
	and bird swarming \cite{eberhart_new_1995,abido_optimal_2002}. In
	the very beginning, PSO is initiated by releasing an initial random
	population of particles into the space. In this paper, the population
	size is fixed. Every particle within the space is a potential solution
	or, in other words, a potential candidate. It is worth mentioning
	that particle and candidate will be used interchangeably. Each single
	particle has a set of parameters such that the set of parameters swarms
	in the space irregularly in a multi-dimensional space. The swarming
	of particles is aimed at determining an optimal solution. The position
	of a particle represents a possible solution including the optimal
	solution. The velocity of every particle has a major impact of heading
	the best potential solution or alternatively, the best candidate position.
	Additionally, the subsequent position and velocity of each particle
	depend on: current velocity, current position, current position of
	the neighboring particles, and best position of the neighboring particles.
	Let $\boldsymbol{\mathcal{P}}\in\mathbb{N}$ and $\boldsymbol{\mathcal{N}}\in\mathbb{N}$
	denote number of parameters to be optimized and population size, respectively.
	For any $i$th candidate solution within the space, define the candidate
	solution (position) at iteration $t\in\mathbb{N}$ by $\mathcal{X}_{i}=\left[x_{i,1}\left(t\right),\ldots,x_{i,\boldsymbol{\mathcal{P}}}\left(t\right)\right]^{\top}\in\mathbb{R}^{\boldsymbol{\mathcal{P}}}$
	and the velocity of every candidate by $\mathcal{V}_{i}=\left[v_{i,1}\left(t\right),\ldots,v_{i,\boldsymbol{\mathcal{P}}}\left(t\right)\right]^{\top}\in\mathbb{R}^{\boldsymbol{\mathcal{P}}}$
	for all $t=1,2,\ldots,\boldsymbol{\mathcal{T}}$ and $i=1,2,\ldots,\boldsymbol{\mathcal{N}}$.
	Accordingly, the velocity update of the $j$th parameter within the
	$i$th swarming particle is given by \cite{eberhart_new_1995,abido_optimal_2002}
	\begin{equation}
	\begin{split}v_{i,j}\left(t\right)= & \alpha\left(t\right)v_{i,j}\left(t-1\right)\\
	& +k_{1}\overline{{\rm rand}}_{1}\left(x_{i,j}^{\star}\left(t-1\right)-x_{i,j}\left(t-1\right)\right)\\
	& +k_{2}\overline{{\rm rand}}_{2}\left(x_{i,j}^{\star\star}\left(t-1\right)-x_{i,j}\left(t-1\right)\right)
	\end{split}
	\label{eq:PSO_v}
	\end{equation}
	where $\alpha\left(t\right)$ denotes an inertial factor to be continuously
	reduced with every iteration such that $\alpha\left(t\right)=0.99\alpha\left(t-1\right)$,
	$\overline{{\rm rand}}_{1}$ and $\overline{{\rm rand}}_{2}$ are
	real random numbers between 0 and 1. $k_{1}$ and $k_{2}$ refer to
	weighting factors associated with personal and social influence, respectively.
	$x_{i,j}^{\star}$ denotes a best local particle within the previous
	generation while $x_{i,j}^{\star\star}$ refers to the best global
	particle within all previous generations, for all $i=1,2,\ldots,\boldsymbol{\mathcal{N}}$
	and $j=1,2,\ldots,\boldsymbol{\mathcal{P}}$. It should be noted that
	the iterative solution starts at $t=1$ continues until the total
	number of iterations set by the user is reached. The position update
	of a parameter within the swarming particle associated with Equation
	\eqref{eq:PSO_v} can be expressed by \cite{eberhart_new_1995,abido_optimal_2002}
	\begin{equation}
	x_{i,j}\left(t\right)=v_{i,j}\left(t\right)+x_{i,j}\left(t-1\right)\label{eq:PSO_p}
	\end{equation}
	The candidate solution $\mathcal{X}_{i}$ is initialized between the
	minimum and maximum boundaries of the space, termed $x_{{\rm min}}$
	and $x_{{\rm max}}$, respectively, such that $x_{i,j}\left(t\right)\in\left[x_{{\rm min}}^{j},x_{{\rm max}}^{j}\right]$.
	On the other side, the velocity $v_{i,j}\left(t\right)$ is initiated
	to be $v_{i,j}\left(t\right)\in\left[-v_{{\rm max}}^{j},v_{{\rm max}}^{j}\right]$,
	while $v_{{\rm max}}$ is given by 
	\begin{equation}
	v_{{\rm max}}=\frac{x_{{\rm max}}-x_{{\rm min}}}{N_{{\rm intv}}}\label{eq:PSO_v_max}
	\end{equation}
	with $N_{{\rm intv}}$ being the number of intervals.
	
	\subsection{Pareto-optimal front and clustering}
	
	For a multi-objective optimization problem, or to be more specific,
	two-objective optimization problem, the trade-off between any two
	solutions can follow two possible scenarios: either one dominates
	the other or none dominates the other. For instance, if $\mathcal{X}_{1}$
	dominates $\mathcal{X}_{2}$ according to the first objective, while
	$\mathcal{X}_{2}$ dominates $\mathcal{X}_{1}$ based on the second
	objective, both solutions are considered non-dominated and are located
	on the Pareto-optimal front. All the solution positioned on the Pareto-optimal
	front are combined into one set. Number of solutions on the Pareto-optimal
	front could be extremely large. Clustering, however, can reduce the
	number of non-dominated solutions within a large set while keeping
	the desired characteristics of the trade-off. In MOPSO, the local
	set of non-dominated solutions is denoted by $\mathcal{S}^{\star}$
	with $\mathcal{S}_{i}^{\star}=\mathcal{X}_{i}\left(1\right)$. After
	which, non-dominated solutions are to be added to form a large set
	$\mathcal{S}_{i}^{\star}$. Clustering is performed based on the distance
	between two pairs. The pairs separated by large distance are retained
	while minimal distance pairs are combined into one cluster \cite{zitzler1999evolutionary}.
	Therefore, the local non-dominated set can be resized to satisfy the
	user settings, for instance $\mathcal{S}_{i}^{\star}$, has the size
	$\mathbf{S}^{{\rm L}}$. The non-dominated global set is denoted by
	$\mathcal{S}^{\star\star}$ and includes all the non-dominated solutions
	from the first to the last generation. In the same spirit, clustering
	is performed to keep pairs within large distance and combine pairs
	within small distance. In that case, the non-dominated global set
	will not exceed a predefined size, for example $\mathcal{S}^{\star\star}$
	has the size $\mathbf{S}^{{\rm G}}$. It should be remarked that $\mathcal{X}_{i}^{\star}$
	and $\mathcal{X}_{k}^{\star\star}$ belong to the sets $\mathcal{S}_{i}^{\star}$
	and $\mathcal{S}^{\star\star}$, respectively. The steps of the clustering
	algorithm can be found in \cite{abido2009multiobjective,abido2010multiobjective}.
	
	Individual distances between all the members of the local set $\mathcal{S}_{i}^{\star}$,
	and the members of the global set $\mathcal{S}^{\star\star}$ are
	measured with respect to the objective space. Let $\mathcal{X}_{i}^{\star}$
	and $\mathcal{X}_{i}^{\star\star}$ denote members of $\mathcal{S}_{i}^{\star}$
	and $\mathcal{S}^{\star\star}$, respectively. For the case when $\mathcal{X}_{i}^{\star}$
	and $\mathcal{X}_{i}^{\star\star}$ give minimum distances, they are
	to be selected as local best and global best of the $i$th particle,
	respectively.
	
	\subsection{Extraction of the best compromise solution}
	
\begin{figure}[!h]
		\centering{}\includegraphics[scale=0.65]{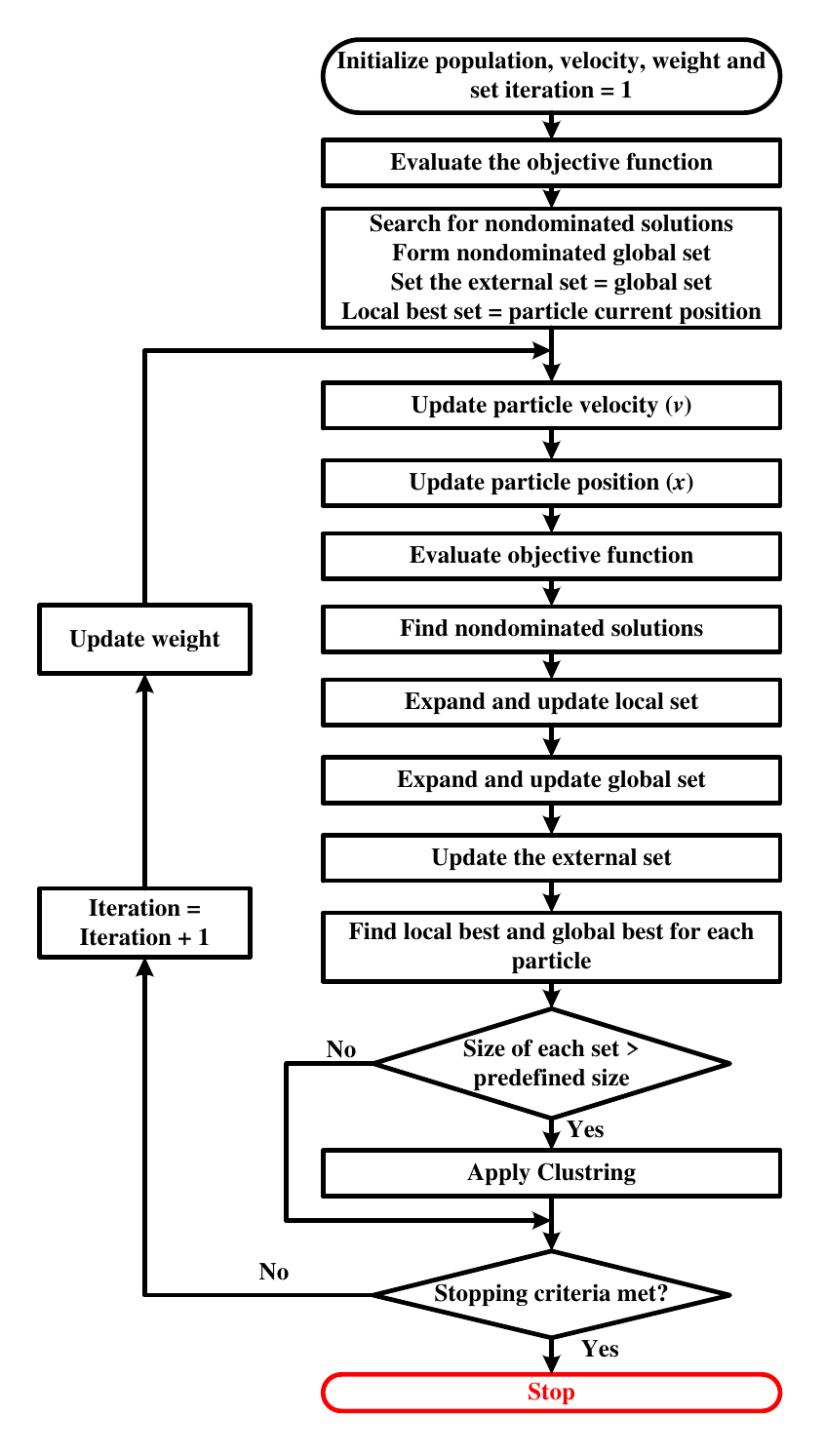} \caption{The computational flowchart of MOPSO \cite{abido2009multiobjective}.}
		\label{Fig_PSO} 
	\end{figure}

Once the Pareto-optimal front is available, one solution should be
selected as the best compromise solution. The fuzzy approach is implemented
to identify the best compromise solution. Define $M$ as a number
of non-dominated objectives on the Pareto-optimal front and $N^{{\rm Obj}}$
as a number of objectives. Let the set of objective functions on the
Pareto-optimal front be $\bar{\boldsymbol{\mathcal{J}}}^{c}=\left[\bar{\boldsymbol{\mathcal{J}}}_{1}^{c},\bar{\boldsymbol{\mathcal{J}}}_{2}^{c},\ldots,\bar{\boldsymbol{\mathcal{J}}}_{M}^{c}\right]\in\mathbb{R}^{M}$
with $\bar{\boldsymbol{\mathcal{J}}}_{{\rm min}}^{c}$ and $\bar{\boldsymbol{\mathcal{J}}}_{{\rm max}}^{c}$
being minimum and maximum objective functions of $\bar{\boldsymbol{\mathcal{J}}}^{c}$,
respectively, for all $c=1,2,\ldots,N^{{\rm Obj}}$ and for $m=1,2,\ldots,M$.
The first step is to obtain the membership function $\mu_{m}^{c}$
such that 
\begin{equation}
\mu_{m}^{c}=\left\{ \begin{array}{cc}
1 & \bar{\boldsymbol{\mathcal{J}}}_{m}^{c}\leq\bar{\boldsymbol{\mathcal{J}}}_{{\rm min}}^{c}\\
\frac{\bar{\boldsymbol{\mathcal{J}}}_{{\rm max}}^{c}-\bar{\boldsymbol{\mathcal{J}}}_{m}^{c}}{\bar{\boldsymbol{\mathcal{J}}}_{{\rm max}}^{c}-\bar{\boldsymbol{\mathcal{J}}}_{{\rm min}}^{c}} & \text{otherwise}\\
0 & \bar{\boldsymbol{\mathcal{J}}}_{m}^{c}\geq\bar{\boldsymbol{\mathcal{J}}}_{{\rm max}}^{c}
\end{array}\right.\label{eq:PSO_Fuzz1}
\end{equation}
for $m=1,2,\ldots,M$, and $c=1,2,\ldots,N^{{\rm Obj}}$. In our case
there are two objectives, thus $N^{{\rm Obj}}=2$. Next, the best
compromise solution is obtained by 
\begin{equation}
\bar{\mu}_{m}=\frac{\sum_{c=1}^{N^{{\rm Obj}}}\mu_{m}^{c}}{\sum_{m=1}^{M}\sum_{c=1}^{N^{{\rm Obj}}}\mu_{m}^{c}}\label{eq:PSO_Fuzz2}
\end{equation}
Given the solution of $\bar{\mu}_{m}$ as in Equation \eqref{eq:PSO_Fuzz2} for $m=1,2,\ldots,M$, the best compromise solution is $\bar{\mu}_{m}$ that has the maximum
value. Figure \ref{Fig_PSO} illustrates the flow chart of MOPSO computational
algorithm.

	\subsection{Implementation of the MOPSO algorithm}
	
	For simplicity, the implementation process of the MOPSO algorithm
	can be subdivided into two stages:
	
	\textbf{Stage I: }Recall Cell/TA-to-TAL/MME Assignment in Section
	\ref{Sec: sysmod}, specifically Equations \eqref{eq: Stage1_eq1},
	\eqref{eq: Stage1_eq2} and \eqref{eq: Stage1_eq3}. There are $\boldsymbol{N}\in\mathbb{N}$
	cells in the system with $\bar{\mathbf{n}}\in\left[1,\boldsymbol{N}\right)$
	cells being distributed within the list and $\bar{\mathbf{n}}<\boldsymbol{N}$:
	\begin{align}
	\bar{\boldsymbol{\mathcal{L}}}^{1} & =\left[\begin{array}{c}
	\mathcal{L}_{1}^{1}\\
	\vdots\\
	\mathcal{L}_{\boldsymbol{N}}^{1}
	\end{array}\right]=\left[\begin{array}{ccc}
	{\rm C}_{1,1}^{1} & \cdots & {\rm C}_{1,\boldsymbol{N}}^{1}\\
	\vdots & \ddots & \vdots\\
	{\rm C}_{\boldsymbol{N},1}^{1} & \cdots & {\rm C}_{\boldsymbol{N},\boldsymbol{N}}^{1}
	\end{array}\right]\in\mathbb{Z}^{\boldsymbol{N}\times\boldsymbol{N}}\nonumber \\
	\vdots\hspace{0.5em} & =\hspace{1em}\hspace{1em}\vdots\hspace{1em}\hspace{1em}\hspace{1em}\hspace{1em}\hspace{1em}\hspace{1em}\hspace{1em}\hspace{1em}\hspace{1em}\hspace{1em}\vdots\nonumber \\
	\bar{\boldsymbol{\mathcal{L}}}^{\boldsymbol{L}} & =\left[\begin{array}{c}
	\mathcal{L}_{1}^{\boldsymbol{L}}\\
	\vdots\\
	\mathcal{L}_{\boldsymbol{N}}^{\boldsymbol{L}}
	\end{array}\right]=\left[\begin{array}{ccc}
	{\rm C}_{1,1}^{\boldsymbol{L}} & \cdots & {\rm C}_{1,\boldsymbol{N}}^{\boldsymbol{L}}\\
	\vdots & \ddots & \vdots\\
	{\rm C}_{\boldsymbol{N},1}^{\boldsymbol{L}} & \cdots & {\rm C}_{\boldsymbol{N},\boldsymbol{N}}^{\boldsymbol{L}}
	\end{array}\right]\in\mathbb{Z}^{\boldsymbol{N}\times\boldsymbol{N}}\label{eq:Simu_Stage1-1}
	\end{align}
	such that $\bar{\boldsymbol{\mathcal{L}}}\in\mathbb{Z}^{\boldsymbol{L}\boldsymbol{N}\times\boldsymbol{N}}$
	denotes all lists and the associated superscript refers to the list
	number. Thus, the number of the cell-to-TAL combination is given by
	$\boldsymbol{L}\times\boldsymbol{N}$. It should be noted that $\bar{\boldsymbol{\mathcal{L}}}^{\boldsymbol{l}}$
	is a symmetric matrix for $\boldsymbol{l}=1,2,\ldots,\boldsymbol{L}$,
	such that $\bar{\boldsymbol{\mathcal{L}}}^{\boldsymbol{l}}=\left(\bar{\boldsymbol{\mathcal{L}}}^{\boldsymbol{l}}\right)^{\top}\in\mathbb{Z}^{\boldsymbol{N}\times\boldsymbol{N}}$,
	with $^{\top}$ being the transpose of the matrix. There are $\bar{\mathbf{n}}$
	nonzero rows within $\bar{\boldsymbol{\mathcal{L}}}^{\boldsymbol{l}}$
	and the rest of rows ($\boldsymbol{N}-\bar{\mathbf{n}}$) are equal
	to zero, while the dimension of each row is $1\times\boldsymbol{N}$.
	Let $\mathcal{L}_{\star_{\boldsymbol{l}}}^{\boldsymbol{l}}\in\mathbb{Z}^{1\times\boldsymbol{N}}$
	be a nonzero row, such that
	$\star_{\boldsymbol{l}}$ is an integer between 1 and $\boldsymbol{N}$ ($\star_{\boldsymbol{l}}\in\mathbb{N}$ and $\star_{\boldsymbol{l}}\in[1,\mathbf{N}]$). In fact, $\bar{\mathbf{n}}$
	numbers within $\mathcal{L}_{\star_{\boldsymbol{l}}}^{\boldsymbol{l}}$
	are equal to 1 and the rest are zeros. Also, $\bar{\mathbf{n}}$ nonzero
	columns within the square matrix $\bar{\boldsymbol{\mathcal{L}}}^{\boldsymbol{l}}$
	are identical, for $\boldsymbol{l}=1,2,\ldots,\boldsymbol{L}$. Accordingly,
	number of variables to be optimized and allocated within one list
	is $\boldsymbol{N}$. In the $\boldsymbol{L}$ lists in Equation \eqref{eq:Simu_Stage1-1},
	there are $\boldsymbol{L}\times\boldsymbol{N}$ binary parameters
	to be optimized and subsequently allocated such that 
	\begin{equation}
	\mathbf{X}^{1}=\left[\begin{array}{cccc}
	\mathcal{L}_{\star_{1}}^{1} & \mathcal{L}_{\star_{2}}^{2} & \cdots & \mathcal{L}_{\star_{L}}^{\boldsymbol{L}}\end{array}\right]^{\top}\in\mathbb{Z}^{\boldsymbol{L}\boldsymbol{N}\times1}\label{eq:Simu_Stage1_X1}
	\end{equation}
	
	\textbf{Stage II:} Consider TAL-to-UE in Section \ref{Sec: sysmod},
	Equation \eqref{eq: Stage2_eq1}. Each TAL is associated with a specific
	MME and their relationship can be expressed as 
	\begin{align}
	\sum\limits _{k=1}^{\boldsymbol{N}}\sigma_{k}^{\boldsymbol{l}}\cdot{\rm C}_{k,n}^{\boldsymbol{l}} & =1,\hspace{1em}\boldsymbol{l}=1,2,\ldots,\boldsymbol{L}\label{eq:Simu_Stage2-1}
	\end{align}
	with $\sigma^{\boldsymbol{l}}=\left[\begin{array}{cccc}
	\sigma_{1}^{\boldsymbol{l}} & \sigma_{2}^{\boldsymbol{l}} & \cdots & \sigma_{\boldsymbol{N}}^{\boldsymbol{l}}\end{array}\right]\in\mathbb{R}^{1\times\boldsymbol{N}}$ for $\boldsymbol{l}=1,2,\ldots,10$. Therefore, in the $\boldsymbol{L}$
	lists in Equation \eqref{eq:Simu_Stage2-1}, there are another $\boldsymbol{L}\times\boldsymbol{N}$
	real parameters to be optimized. The decision variable $\sigma^{\boldsymbol{l}}$
	has to match the $\mathcal{L}_{\star_{\boldsymbol{l}}}^{\boldsymbol{l}}$,
	for instance
	\begin{equation}
	\sigma_{k}^{\boldsymbol{l}}=\begin{cases}
	0, & \text{if }{\rm C}_{\star_{\boldsymbol{l}},k}^{\boldsymbol{l}}=0\\
	\sigma_{k}^{\boldsymbol{l}}, & \text{if }{\rm C}_{\star_{\boldsymbol{l}},k}^{\boldsymbol{l}}=1
	\end{cases}\label{eq:Simu_Stage2-Condition1-1}
	\end{equation}
	In order to obtain $\sigma^{\boldsymbol{l}}$, define the following
	auxiliary variable $\bar{\sigma}^{\boldsymbol{l}}=\left[\begin{array}{cccc}
	\bar{\sigma}_{1}^{\boldsymbol{l}} & \bar{\sigma}_{2}^{\boldsymbol{l}} & \cdots & \bar{\sigma}_{\boldsymbol{N}}^{\boldsymbol{l}}\end{array}\right]\in\mathbb{R}^{1\times\boldsymbol{N}}$ with $\bar{\sigma}_{k}^{\boldsymbol{l}}\in\left[0,1\right]$ for
	all $\boldsymbol{l}=1,2,\ldots,10$ and $k=1,2,\ldots,\boldsymbol{N}$.
	Hence, we optimize for $\mathbf{X}^{2}$ such that 
	\begin{equation}
	\mathbf{X}^{2}=\left[\begin{array}{cccc}
	\bar{\sigma}^{1} & \bar{\sigma}^{2} & \cdots & \bar{\sigma}^{\boldsymbol{N}}\end{array}\right]^{\top}\in\mathbb{R}^{\boldsymbol{L}\boldsymbol{N}\times1}\label{eq:Simu_Stage2_X2}
	\end{equation}
	After optimizing $\mathbf{X}^{2},$ $\sigma^{\boldsymbol{l}}$ can
	be defined as follows 
	\begin{equation}
	\sigma_{k}^{\boldsymbol{l}}=\begin{cases}
	0, & \text{if }{\rm C}_{\star_{\boldsymbol{l}},k}^{\boldsymbol{l}}=0\\
	\bar{\sigma}_{k}^{\boldsymbol{l}}, & \text{if }{\rm C}_{\star_{\boldsymbol{l}},k}^{\boldsymbol{l}}=1
	\end{cases}\label{eq:Simu_Stage2-Condition1}
	\end{equation}
	where ${\rm C}_{\star_{\boldsymbol{l}},k}^{1}$ is a cell associated
	with the nonzero row $\mathcal{L}_{\star_{\boldsymbol{l}}}^{\boldsymbol{l}}$.
	Next, to satisfy the equality in Equation \eqref{eq:Simu_Stage2-1},
	one has 
	\begin{equation}
	\sigma^{\boldsymbol{l}}=\frac{\sigma^{\boldsymbol{l}}}{\sum\limits _{i=1}^{\boldsymbol{N}}\sigma_{k}^{\boldsymbol{l}}}\label{eq:Simu_Stage2-Condition2}
	\end{equation}
	This concludes the implementation process of the MOPSO algorithm.
	
	Therefore, \textbf{Stage I }and \textbf{Stage II} summarize the optimization
	process and number of parameters to be optimized. It is worth re-mentioning
	that $\mathbf{X}^{1}$ and $\mathbf{X}^{2}$ in\textbf{ Stage I }and
	\textbf{Stage II}, respectively, would be solved by MOPSO algorithm
	simultaneously. In total, there are $2\boldsymbol{L}\boldsymbol{N}$
	parameters to be optimized. Let $\mathcal{X}_{i}$ denote the vector
	of parameters to be optimized with dimension $2\boldsymbol{L}\boldsymbol{N}\times1$
	for all $i=1,2,\ldots,\boldsymbol{\mathcal{N}}$. One can find the
	complete vector ($\mathcal{X}_{i}$) to be 
	\begin{equation}
	\mathcal{X}_{i}=\left[\begin{array}{c}
	\mathbf{X}^{1}\\
	\mathbf{X}^{2}
	\end{array}\right]=\left[x_{i,1},\ldots,x_{i,\boldsymbol{\mathcal{P}}}\right]^{\top}\label{eq:Simu_Stage12_X}
	\end{equation}
	The total number of parameters to be optimized in one particle is
	$\boldsymbol{\mathcal{P}}=2\boldsymbol{L}\boldsymbol{N}$ with $\mathbf{X}^{1}\in\mathbb{Z}^{\boldsymbol{N}\times1}$
	and $\mathbf{X}^{2}\in\mathbb{R}^{\boldsymbol{N}\times1}$.
	
	A set of random particles is initiated such that every single particle
	has $\boldsymbol{\mathcal{P}}=2\boldsymbol{L}\boldsymbol{N}$ parameters,
	and there are $\boldsymbol{\mathcal{N}}$ particles within one iteration/generation.
	Therefore, the number of objective functions obtained from one generation
	is $2\boldsymbol{L}\boldsymbol{N}$. Velocity and position of the
	$i$th particle are defied according to Equations \eqref{eq:PSO_v}
	and \eqref{eq:PSO_p}, respectively. All the remaining steps are detailed
	above.
	
	\subsection{Computational flow using MOPSO }
	
	The computational flow of the proposed location management of LTE
	networks using MOPSO technique can be briefly summarized as follows:
	
	\textbf{Step 1} (Initialization): initiate $t=1$ and generate $\boldsymbol{\mathcal{N}}$
	random particles $\mathcal{X}_{i}\left(1\right)\in\mathbb{R}^{\boldsymbol{\mathcal{P}}}\forall i=1,2,\ldots,\boldsymbol{\mathcal{N}}$
	where $\mathcal{X}_{i}\left(1\right)=\left[\mathbf{X}^{1^{\top}},\mathbf{X}^{2^{\top}}\right]^{\top}=\left[x_{i,1}\left(1\right),\ldots,x_{i,\boldsymbol{\mathcal{P}}}\left(1\right)\right]^{\top}\in\mathbb{R}^{\boldsymbol{\mathcal{P}}}$
	such that $\mathbf{X}^{1}\in\mathbb{Z}^{\boldsymbol{N}\times1}$,
	$\mathbf{X}^{2}\in\mathbb{R}^{\boldsymbol{N}\times1}$, and $x_{i,k}\left(1\right)\in\left[x_{{\rm min}}^{k},x_{{\rm max}}^{k}\right]$.
	Likewise, generate $\boldsymbol{\mathcal{N}}$ random initial velocities
	$\mathcal{V}_{i}\left(1\right)\in\mathbb{R}^{\boldsymbol{\mathcal{P}}}\forall i=1,2,\ldots,\boldsymbol{\mathcal{N}}$
	where $\mathcal{V}_{i}\left(1\right)=\left[v_{i,1}\left(1\right),\ldots,v_{i,\boldsymbol{\mathcal{P}}}\left(1\right)\right]^{\top}$
	and $v_{i,k}\left(1\right)\in\left[-v_{{\rm max}}^{k},v_{{\rm max}}^{k}\right]$.
	Evaluate the objective function of every particle. For every particle,
	generate a local best set $\mathcal{S}_{i}^{\star}\left(1\right)$
	with $\mathcal{X}_{i}^{\star}\left(1\right)=\mathcal{X}_{i}\left(1\right)$
	and $\mathcal{S}_{i}^{\star}\left(1\right)=\mathcal{X}_{i}^{\star}\left(1\right)$
	for all $i=1,2,\ldots,\boldsymbol{\mathcal{N}}$. Search for non-dominated
	solutions and generate a non-dominated global set $\mathcal{S}^{\star\star}\left(1\right)$
	where the closest member of $\mathcal{S}^{\star\star}\left(1\right)$
	to $\mathcal{X}_{i}^{\star}\left(1\right)$ is marked as the global
	best $\mathcal{X}_{i}^{\star\star}\left(1\right)$ of the $i$th particle.
	Set the initial value of the inertia factor $\alpha\left(1\right)$.
	
	\textbf{Step 2} (iteration update): update the iteration number to
	$t=t+1$.
	
	\textbf{Step 3} (Inertia factor update): update the inertia factor.
	
	\textbf{Step 4} (Velocity update): update the velocity $\mathcal{V}_{i}\left(t\right)$
	as defined in Equation \eqref{eq:PSO_v} given the particle $\mathcal{X}_{i}\left(t-1\right)$,
	the local best $\mathcal{X}_{i}^{\star}\left(t-1\right)$, and the
	global best $\mathcal{X}_{i}^{\star\star}\left(t-1\right)$ for all
	$i=1,2,\ldots,\boldsymbol{\mathcal{N}}$.
	
	\textbf{Step 5} (Position update): update the position $\mathcal{X}_{i}\left(t\right)$
	in accordance with Equation \eqref{eq:PSO_p} given the particle $\mathcal{X}_{i}\left(t-1\right)$
	and the velocity $\mathcal{V}_{i}\left(t\right)$ for all $i=1,2,\ldots,\boldsymbol{\mathcal{N}}$.
	
	\textbf{Step 6} (Non-dominated local set update): The updated position
	$\mathcal{X}_{i}\left(t\right)$ is added to the set $\mathcal{S}_{i}^{\star}\left(t\right)$
	for all $i=1,2,\ldots,\boldsymbol{\mathcal{N}}$. Keep all non-dominated
	solutions and eliminate all dominated solutions within the local best
	set $\mathcal{S}_{i}^{\star}\left(t\right)$. If the size of $\mathcal{S}_{i}^{\star}\left(t\right)$
	exceeds the predefined value $\mathbf{S}^{{\rm L}}$, apply clustering
	and reduce the size to $\mathbf{S}^{{\rm L}}$.
	
	\textbf{Step 7} (Non-dominated global set update): Set $\mathcal{S}^{\star\star}\left(t\right)=\mathcal{S}^{\star\star}\left(t-1\right)$.
	If any of the non-dominated solutions within $\mathcal{S}_{i}^{\star}\left(t\right)$
	are non-dominated by $\mathcal{S}^{\star\star}\left(t\right)$ and
	are not its members, add it to the set $\mathcal{S}^{\star\star}\left(t\right)$.
	If the size of $\mathcal{S}^{\star\star}\left(t\right)$ exceeds the
	predefined limit $\mathbf{S}^{{\rm G}}$, apply clustering to reduce
	the size to $\mathbf{S}^{{\rm G}}$.
	
	\textbf{Step 8} (Local best and global best update): The individual
	distances between members of $\mathcal{S}_{i}^{\star}$ and members
	of $\mathcal{S}^{\star\star}\left(t\right)$ are evaluated with respect
	to the space of the objective function. The pair that gives minimum
	distance is selected as local best and global best of the $i$th particle,
	respectively.
	
	\textbf{Step 9} (Stopping criteria): If the maximum number of iterations
	defined by the user is reached, stop the program. Otherwise, go to
	\textbf{Step 2}. 
	
	\section{Performance Evaluation and Simulations \label{Sec: 5 Simulation} }
	
	Optimization of the LTE networks location management using MOPSO is
	evaluated and compared to MINLP using ${\rm MATLAB}{}^{\circledR}$.
	In order to achieve reliable assessment, both algorithms, MOPSO and
	MINLP, have been implemented at different speed levels. The speed
	range has been subdivided into four levels: very slow speed stands
	for (0-8 m/s); slow speed indicated (8-16 m/s); normal speed should
	be interpreted as (16-25 m/s); and high speed implies speeds in the
	(25-33 m/s) range. The environment is assumed to have 30 cells with
	an average of 100 users are evenly distributed throughout the systems
	of cells. The users are allocated in a regular manner through the
	cells within the system. The system consists of 10 lists with 0.05
	rate of paging. Clearly, the number of lists in the system equals
	the total number of MME. Table \ref{tab: Table3} outlines the full
	details of the model parameters associated with LTE networks.
	
	\begin{table}[H]
		\centering{}\caption{\label{tab: Table3}Simulation Parameters \& Values}
		\begin{tabular}{ll}
			\toprule 
			\textbf{Parameter}  & \textbf{Value} \tabularnewline
			\midrule
			\midrule 
			Total number of TAs  & 30\tabularnewline
			\midrule 
			Cells number ($\boldsymbol{N}$)  & 30\tabularnewline
			\midrule 
			Total number of lists ($\boldsymbol{L}$)  & 10\tabularnewline
			\midrule 
			Users average number  & 100 per cell $k$ \tabularnewline
			\midrule 
			Paging rate ($\mathbf{G}^{{\rm a}}$)  & 0.05 \tabularnewline
			\midrule 
			Number of UE speeds  & 4\tabularnewline
			\midrule 
			UE speeds  & 0, 8, 16, 25 and 33 (m/sec) \tabularnewline
			\midrule 
			Radius of the cell  & 500 m\tabularnewline
			\bottomrule
		\end{tabular}
	\end{table}

	\subsection{MOPSO Set-up}
	
	The aim of the simulation section is to examine the convergence of
	MOPSO in comparison with MINLP. Recall \textbf{Stage I} subsection
	of the Cell/TA-to-TAL/MME Assignment (Section \ref{Sec: 4 MO-PSO}),
	more specifically, Equations \eqref{eq:Simu_Stage1-1}, \eqref{eq: Stage1_eq2}
	and \eqref{eq: Stage1_eq3}. As mentioned above, there are $\boldsymbol{N}=30$
	cells in the system with $\bar{\mathbf{n}}=16$ cells being distributed
	within the list. Thus, $\bar{\boldsymbol{\mathcal{L}}}^{\boldsymbol{l}}\in\mathbb{Z}^{30\times30}$
	denotes a complete list. Define $\mathcal{L}_{\star_{\boldsymbol{l}}}^{\boldsymbol{l}}\in\mathbb{Z}^{30\times1}$
	to be a nonzero row, such that $\star_{\boldsymbol{l}}$
	is any integer between 1 and 30. $\mathcal{L}_{\star_{\boldsymbol{l}}}^{\boldsymbol{l}}$
	includes 16 cells with the value of 1 while the rest of the cells
	have the value of 0. For $\boldsymbol{L}=10$ lists in Equation \eqref{eq:Simu_Stage1-1},
	number of parameters to be optimized is 300 such that 
	\[
	\mathbf{X}^{1}=\left[\begin{array}{cccc}
	\mathcal{L}_{\star_{1}}^{1} & \mathcal{L}_{\star_{2}}^{2} & \cdots & \mathcal{L}_{\star_{10}}^{10}\end{array}\right]^{\top}\in\mathbb{Z}^{300\times1}
	\]
	Each TAL is related to MME and the relation can be expressed by 
	\begin{align}
	\sum\limits _{k=1}^{30}\sigma_{k}^{\boldsymbol{l}}\cdot{\rm C}_{k,n}^{\boldsymbol{l}} & =1,\hspace{1em}\boldsymbol{l}=1,2,\ldots,10\label{eq:Simu_Stage2}
	\end{align}
	with $\sigma^{\boldsymbol{l}}=\left[\begin{array}{cccc}
	\sigma_{1}^{\boldsymbol{l}} & \sigma_{2}^{\boldsymbol{l}} & \cdots & \sigma_{30}^{\boldsymbol{l}}\end{array}\right]\in\mathbb{R}^{1\times30}$ for $\boldsymbol{l}=1,2,\ldots,10$. A new variable $\bar{\sigma}^{\boldsymbol{l}}$
	is introduced such that $\bar{\sigma}^{\boldsymbol{l}}=\left[\begin{array}{cccc}
	\bar{\sigma}_{1}^{\boldsymbol{l}} & \bar{\sigma}_{2}^{\boldsymbol{l}} & \cdots & \bar{\sigma}_{30}^{\boldsymbol{l}}\end{array}\right]\in\mathbb{R}^{1\times30}$. For 10 lists in Equation \eqref{eq:Simu_Stage2-1}, one has 300
	real parameters to be optimized, such that 
	\[
	\mathbf{X}^{2}=\left[\begin{array}{cccc}
	\bar{\sigma}^{1} & \bar{\sigma}^{2} & \cdots & \bar{\sigma}^{10}\end{array}\right]^{\top}\in\mathbb{R}^{300\times1}
	\]
	therefore, the decision $\sigma^{\boldsymbol{l}}$ is defined by 
	\[
	\sigma_{k}^{\boldsymbol{l}}=\begin{cases}
	0, & \text{if }{\rm C}_{\star_{\boldsymbol{l}},k}^{\boldsymbol{l}}=0\\
	\bar{\sigma}_{k}^{\boldsymbol{l}}, & \text{if }{\rm C}_{\star_{\boldsymbol{l}},k}^{\boldsymbol{l}}=1
	\end{cases}
	\]
	next, one has
	\[
	\sigma^{\boldsymbol{l}}=\frac{\sigma^{\boldsymbol{l}}}{\sum\limits _{k=1}^{30}\sigma_{k}^{\boldsymbol{l}}}
	\]
	Therefore, there are 600 parameters to be optimized. Define vector
	$\mathcal{X}_{i}$ as the total number of parameters to be optimized
	such that 
	\[
	\mathcal{X}_{i}=\left[\begin{array}{c}
	\mathbf{X}^{1}\\
	\mathbf{X}^{2}
	\end{array}\right],\hspace{1em}\forall\mathbf{X}^{1}\in\mathbb{Z}^{300\times1}\text{ and }\mathbf{X}^{2}\in\mathbb{R}^{300\times1}
	\]
	And the problem of optimization is defined as a minimization problem
	using MOPSO. The setting parameters of MOPSO are listed in Table \ref{Tab_Settings}
	with $\boldsymbol{\mathcal{T}}=400$ being the total number of iterations.
	Also, it should be noted that the setting of parameters in Table \ref{Tab_Settings}
	were selected as a result of a set of trials.
	
	\begin{table}[H]
		\caption{\label{Tab_Settings} Setting Parameters of MOPSO algorithm}
		
		\centering{}%
		\begin{tabular}{cccccccc}
			\toprule 
			Parameter  & $\boldsymbol{\mathcal{P}}$  & $\boldsymbol{\mathcal{N}}$  & $N_{{\rm intv}}$  & $k_{1}$  & $k_{2}$  & $\mathbf{S}^{{\rm L}}$  & $\mathbf{S}^{{\rm G}}$ \tabularnewline
			\midrule
			Setting  & 600  & 10000  & 5 & 2 & 2 & 5 & 10\tabularnewline
			\bottomrule
		\end{tabular}
	\end{table}
	The inertia factor $\alpha\left(t\right)$ is calculated by $\alpha\left(t\right)=0.99\alpha\left(t-1\right)$
	and initiated at $\alpha\left(1\right)=1$.
	
	\subsection{MOPSO Total Signaling Cost}
	
\begin{figure*}[h!]
	\centering{}\includegraphics[scale=0.41]{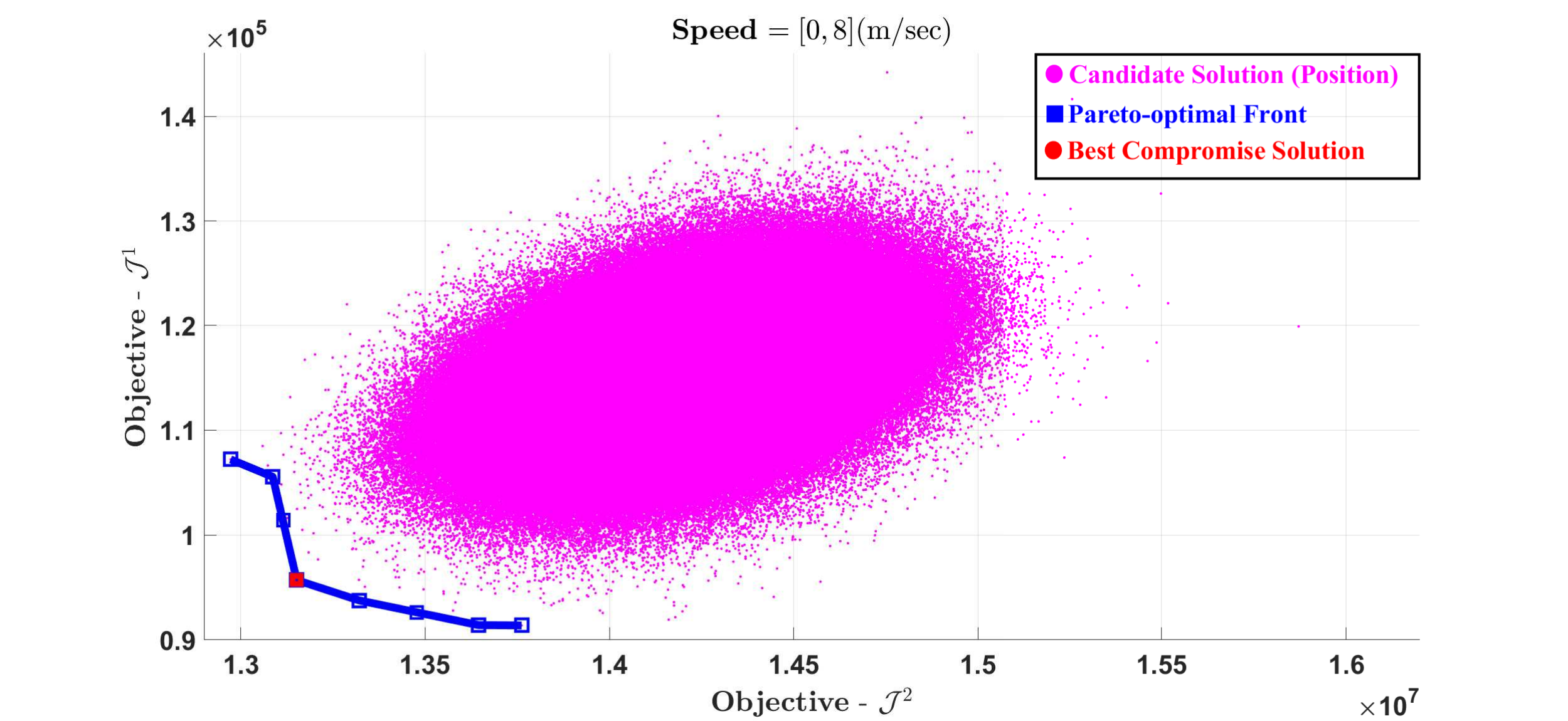}\caption{Objective function minimization of speed = {[}0, 8{]} (m/sec): Search history, Pareto-optimal front and best compromise solution. }
	\label{Fig_Obj_all} 
\end{figure*}

\begin{figure}[h!]
	\centering{}\includegraphics[scale=0.23]{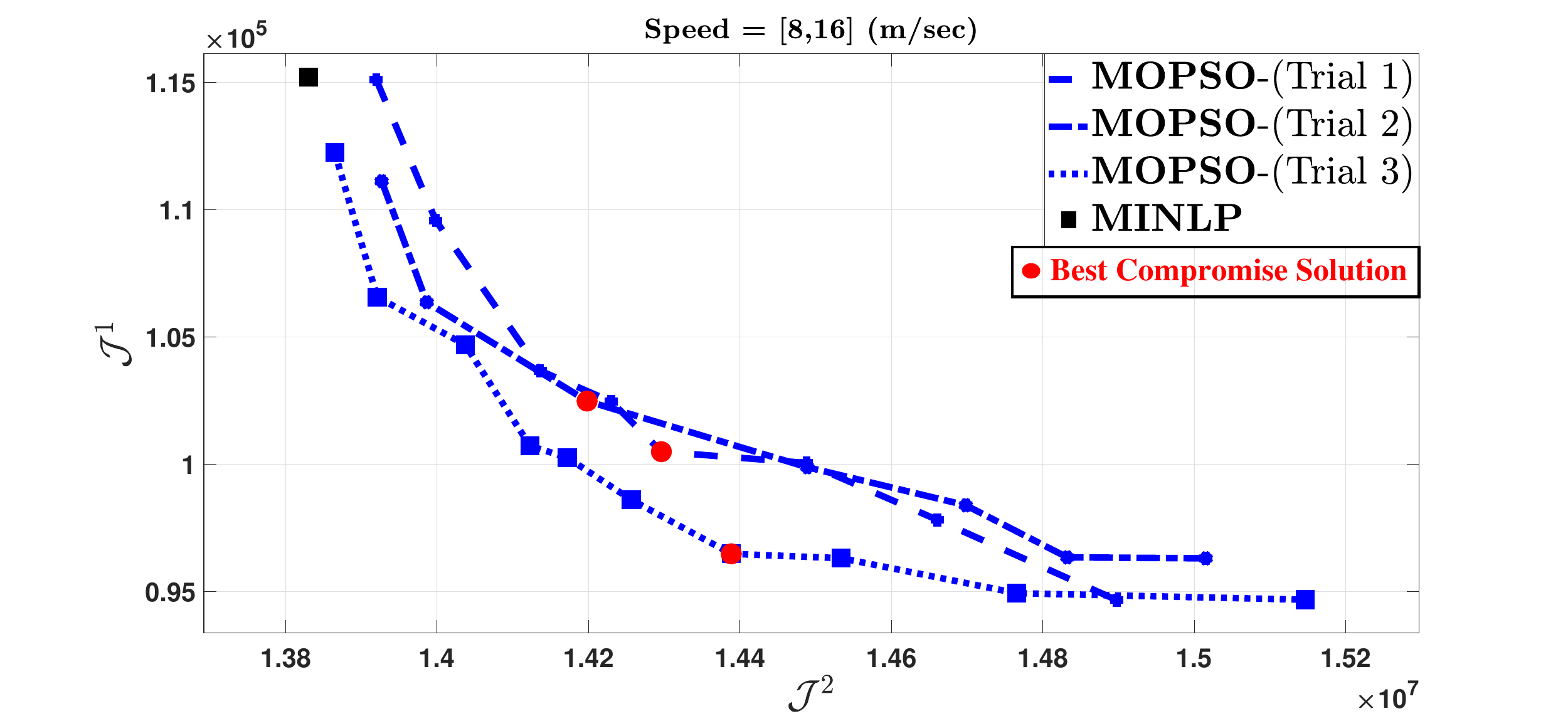}\caption{Objective function minimization of speed = {[}8, 16{]} (m/sec): MOPSO
		vs MINLP.}
	\label{Fig_Obj1} 
\end{figure}

\begin{figure}[h!]
	\centering{}\includegraphics[scale=0.23]{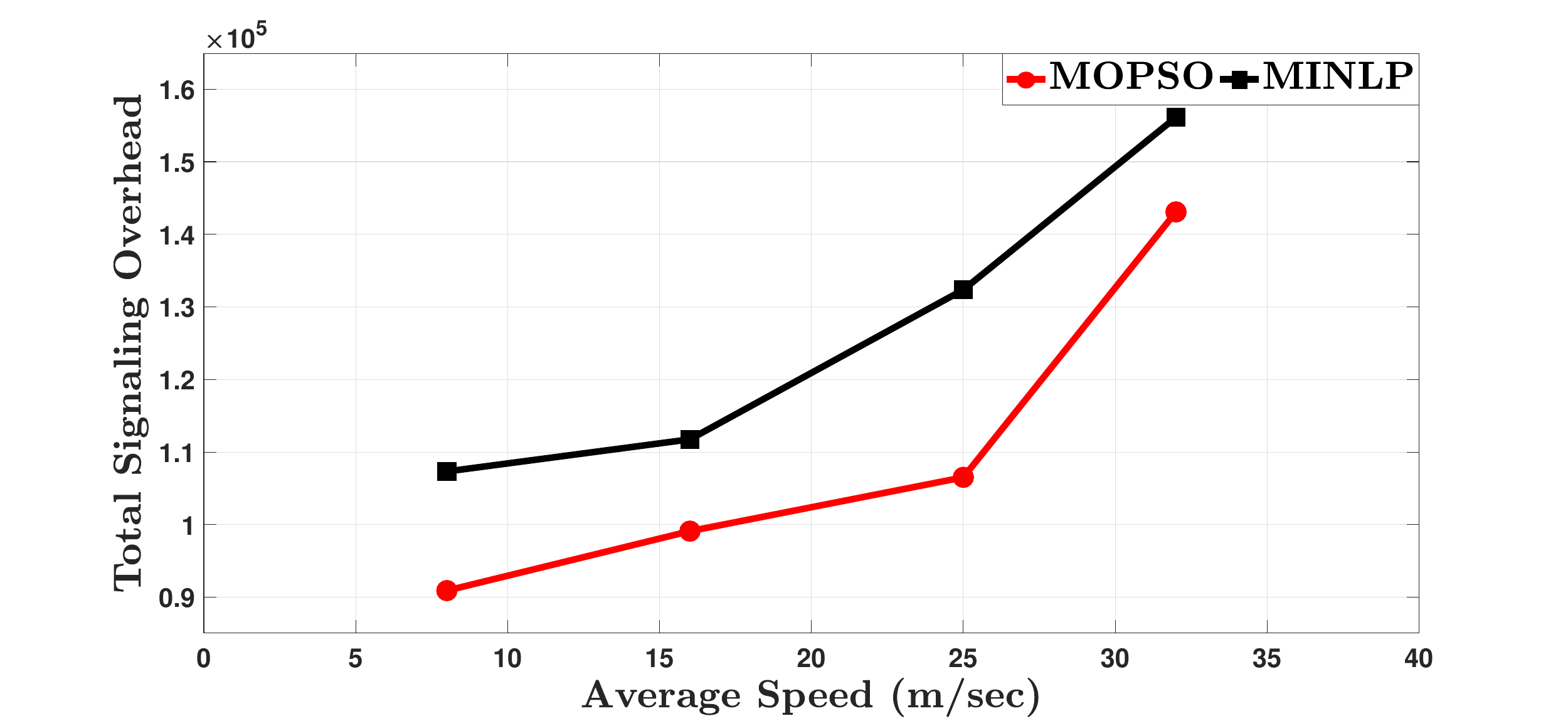}\caption{Average total signaling overhead: MOPSO vs MINLP.}
	\label{Fig_Total_Signaling} 
\end{figure}

\begin{figure}[h!]
	\centering{}\includegraphics[scale=0.23]{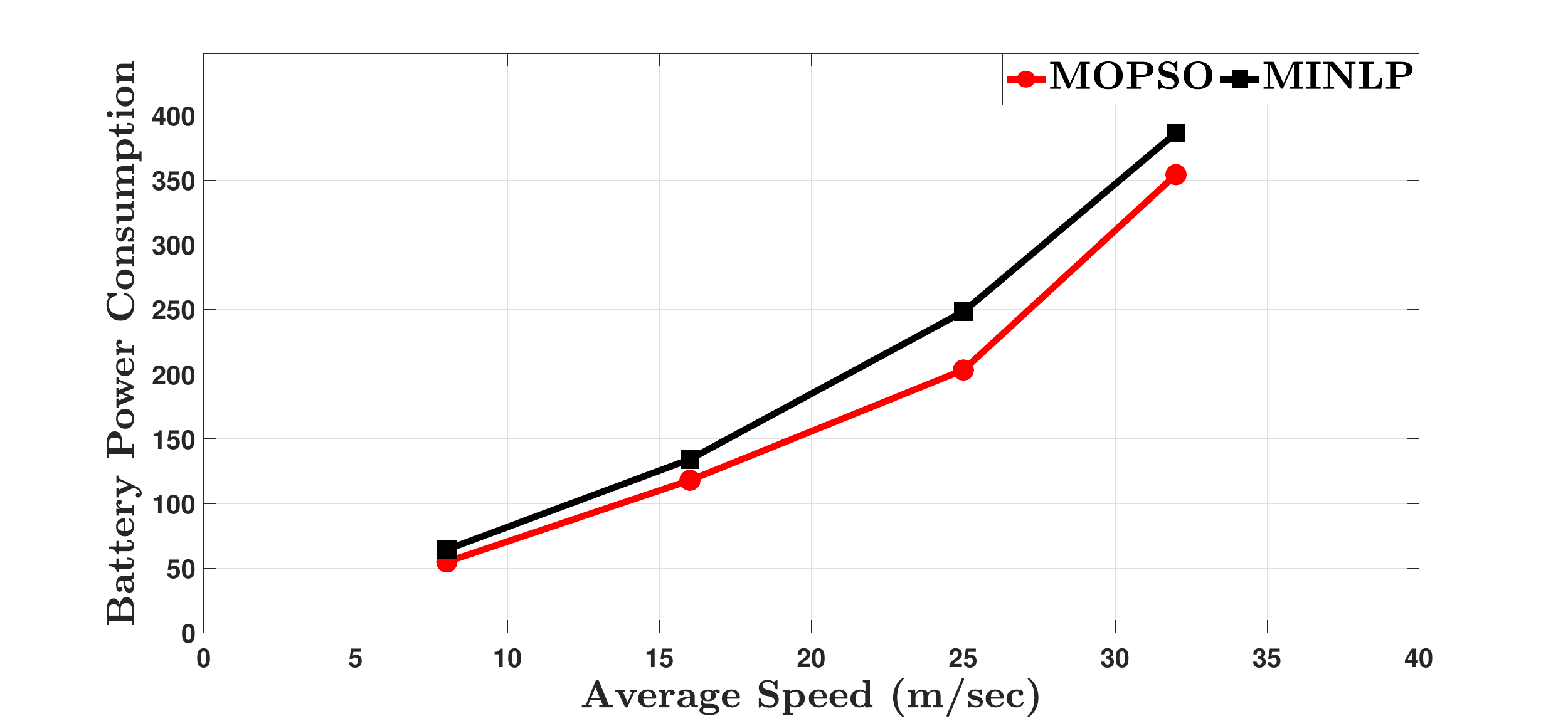}\caption{Average battery power consumption: MOPSO vs MINLP.}
	\label{Fig_Power} 
\end{figure}

	Using MOPSO, the total signaling of TAU and paging ($\boldsymbol{\mathcal{J}}^{1}$)
	and the total inter-list handover ($\boldsymbol{\mathcal{J}}^{2}$)
	are presented and compared to MINLP at four different speeds mentioned
	above. The optimization algorithms, MOPSO and MINLP have been implemented
	and compared to assess: 
	\begin{enumerate}
		\item The minimum value of the two objective functions $\boldsymbol{\mathcal{J}}^{1}$
		and $\boldsymbol{\mathcal{J}}^{2}$ obtained by each algorithm. 
		\item The best compromise solution by MOPSO and the MINLP solution.
		\item The totaling signaling overhead and average battery power consumption.
	\end{enumerate}
	In order to evaluate the robustness of MOPSO algorithm in terms of
	its ability to approach the optimal solution, the algorithm has been
	implemented four times at each of the four speed levels considering
	different random initialization conditions of the $\mathcal{X}$ vector.
	The speed ranges examined are the following: {[}0,8{]}, {[}8,16{]},
	{[}16,25{]} and {[}25,33{]} m/sec. For clarity, three out of the four
	trials have been presented. Additionally, the following color notation
	is used: red color demonstrates a best compromise solution, blue illustrates
	a solution on the Pareto-optimal front and magenta represents a candidate
	solution as well as search history obtained using MOPSO. Also, black
	refers to a solution obtained by MINLP. An illustration of the search
	history of a single trial of the speed {[}0,8{]} (m/sec) within the
	two-dimensional space is depicted in Figure \ref{Fig_Obj_all} presenting
	all candidate solutions, Pareto-optimal front and the best compromise
	solution. Figure \ref{Fig_Obj1} %
	demonstrate the Pareto-optimal front in blue color and the best compromise
	solutions in red color obtained by MOPSO plotted against MINLP in
	black color at speed ranges {[}8,16{]}%
	. In spite of the fact that the solution generated by MINLP is located
	on the Pareto-optimal front as depicted in Figure \ref{Fig_Obj1}%
	, MINLP favored the objective function $\boldsymbol{\mathcal{J}}^{2}$
	over $\boldsymbol{\mathcal{J}}^{1}$. MOPSO, on the contrary, captured
	a set of solutions on the Pareto-optimal front with the trade-off
	represented by the best compromise solution. The best compromise solution
	creates a balance between the inter-list handover which resides on
	UEs, probability of the user to travel between cells, and the cost
	of inter-MME, from one side, and the total signaling cost of TAU and
	paging which impact the EPC significantly by influencing the location
	management of UEs, on the other side. The significance of the proposed
	algorithm can be observed in terms of total signaling overhead and
	battery power consumption. The consumption estimation of the triggered
	TAU signal in a regular smart-phone is considered 10 mW \cite{razavi2014reducing}.
	Figure \ref{Fig_Total_Signaling} illustrates the total signaling
	overhead of MOPSO best compromise solution average values versus MINLP.
	MOPSO exhibited lower total signaling overhead than MINLP at various
	speed ranges as presented in Figure \ref{Fig_Total_Signaling}. As
	Figure \ref{Fig_Power} confirms that MOPSO has a significant advantage
	over MINLP with respect to battery power consumption. The average
	values of MOPSO best compromise solutions associated with battery
	power consumption versus MINLP are depicted in Figure \ref{Fig_Power}.
	
	The overall quality of MOPSO convergence very close to the optimal
	solution is presented in terms of mean and standard deviation (STD).
	Table \ref{Tab_Obj} lists the best compromise solutions of non-dominated
	objective functions recorded on the Pareto-optimal front considering
	four trials at various speed ranges. Also, Table \ref{Tab_Obj} outlines
	the mean and STD of the four trials at every speed range. The accuracy
	of convergence can be evaluated by the percentage of STD with respect
	to the means, which is known as a relative standard deviation (RSD).
	RSD of $\boldsymbol{\mathcal{J}}^{1}$ is 0.7\%, 0.7\%, 1.3\% and
	1.26\% and RSD of $\boldsymbol{\mathcal{J}}^{2}$ is 1.4\%, 2.8\%,
	1.2\% and 0.68\% at the four speed ranges. Indeed, the RSD of $\boldsymbol{\mathcal{J}}^{1}$
	and $\boldsymbol{\mathcal{J}}^{2}$ illustrates the remarkable performance
	of MOPSO when applied to the LTE networks location management. Likewise,
	the total power consumption has been evaluated in Table \ref{Tab_Power}
	in terms of the mean and STD of the results obtained by MOPSO algorithm
	versus MINLP at different speeds. It can be noticed that MOPSO generated
	lower values of total power consumption than MINLP.
	
	\begin{table*}[!h]
		\centering{}\caption{\label{Tab_Obj}MOPSO best compromise solution of the four trials
			at various range of speeds}
		\begin{tabular}{c|c|c|c|c|c}
			\hline 
			\multicolumn{1}{c|}{Experiment index} & 1  & 2  & 3  & 4  & Mean $\pm$ STD\tabularnewline
			\hline 
			\multicolumn{1}{c|}{Speed Range} & \multicolumn{5}{c}{{[}0-8{]} m/sec }\tabularnewline
			\hline 
			\hline 
			$\boldsymbol{\mathcal{J}}^{1}$ & 94343 & 91091 & 92738 & 93432 & 92901$\pm$ 1374 \tabularnewline
			$\boldsymbol{\mathcal{J}}^{2}$ & 13342339 & 13325501 & 13436437 & 13204385 & 13327165$\pm$ 92901 \tabularnewline
			\hline 
			\multicolumn{1}{c|}{Speed Range} & \multicolumn{5}{c}{{[}8-16{]} m/sec }\tabularnewline
			\hline 
			\hline 
			$\boldsymbol{\mathcal{J}}^{1}$ & 97236 & 100491 & 102482 & 96479 & 99172$\pm$ 2810 \tabularnewline
			$\boldsymbol{\mathcal{J}}^{2}$ & 14161850 & 14296212 & 14198106 & 14388572 & 14261185$\pm$ 102144 \tabularnewline
			\hline 
			\multicolumn{1}{c|}{Speed Range} & \multicolumn{5}{c}{{[}16-25{]} m/sec }\tabularnewline
			\hline 
			\hline 
			$\boldsymbol{\mathcal{J}}^{1}$ & 110289 & 107139 & 108643 & 108418 & 108622$\pm$ 1293 \tabularnewline
			$\boldsymbol{\mathcal{J}}^{2}$ & 16045610 & 16512041 & 16202005 & 16431714 & 16297842$\pm$ 213392 \tabularnewline
			\hline 
			\multicolumn{1}{c|}{Speed Range} & \multicolumn{5}{c}{{[}25-33{]} m/sec }\tabularnewline
			\hline 
			\hline 
			$\boldsymbol{\mathcal{J}}^{1}$ & 145550 & 143241 & 144100 & 144786 & 144419$\pm$ 983 \tabularnewline
			$\boldsymbol{\mathcal{J}}^{2}$ & 18502495 & 18567046 & 18509153 & 18115217 & 18423477$\pm$ 207541 \tabularnewline
			\hline 
		\end{tabular}
	\end{table*}

	\begin{table}
		\centering{}\caption{\label{Tab_Power} Total power consumption Mean and STD of MOPSO against
			MINLP at various speed ranges}
		\begin{tabular}{c|c|c|c|c}
			\hline 
			\multicolumn{1}{c|}{Speed range} & 0 - 8  & 8 - 16  & 16 - 25  & 25 - 33 \tabularnewline
			\hline 
			Algorithm  & \multicolumn{4}{c}{Mean $\pm$ STD}\tabularnewline
			\hline 
			\hline 
			MOPSO  & 54 $\pm$ 0.82  & 117 $\pm$ 2.7  & 203 $\pm$ 3.8  & 354 $\pm$ 2.89 \tabularnewline
			\hline 
			MINLP & 64.5  & 134  & 248 & 386.5\tabularnewline
			\hline 
		\end{tabular}
	\end{table}

	\section{Conclusion \textcolor{blue}{\label{Sec: conclusion} }}
	
	Cellular networks are essential as they offer a variety of connectivity
	solutions. LTE and LTE-A provide reliable and fast wireless network
	service. Nonetheless, the total signaling overhead has been a concern,
	in particular with the rapid growth of more advanced cellular phones
	and phone applications. In fact, location management which allocate
	the idle user is directly related to the total signaling overhead.
	The total signaling overhead can be assessed by two elements, the
	total signaling cost of TAU and paging and the total inter-list handover.
	However, these two elements are adversely related. This study proposes
	a new formulation of the total signaling overhead as a true multi-objective
	problem where both conflicting objectives are optimized simultaneously.
	This novel approach is based on the multi-objective particle swarm
	optimization (MOPSO) which allows to minimize the total signaling
	overhead, and therefore bring location management in LTE networks
	to a qualitatively new level. The significant decrease in the total
	signaling overhead is achieved by minimizing the two objectives and
	obtaining a best compromise solution from a set of non-dominated solutions
	on the Pareto-optimal front. The first objective is the lessening
	of the total signaling cost of TAU and paging and the second objective
	is the minimization of the total inter-list handover. A set of experiments
	has been performed considering different random initialization to
	validate the robustness of the proposed algorithm. Location management
	in LTE networks using MOPSO has been examined by considering a large
	scale environment problem. In this problem various speed ranges have
	been considered with four experiments at every speed range. The best
	compromise solution obtained by MOPSO is compared to MINLP. Lower
	values of total signaling overhead and power consumption have been
	observed using MOPSO than MINLP. In addition, at every speed range,
	small values of relative standard deviation (RSD) have been observed.
	The robustness of MOPSO algorithm in terms of its ability to approach
	the optimal solution has been validated through a set of four algorithm
	runs performed at each level of the four different speed levels considering
	different random initialization conditions of the optimized parameters.
	Thus the proposed algorithm adequately represents a network with a
	variety of mobility patterns and different cells. \textcolor{blue}{}%

\section*{Acknowledgment}
The authors would like to thank \textbf{Maria Shaposhnikova} for proofreading the article. Also, Dr. Abido would like to acknowledge the funding support provided by King Abdullah City for Atomic and Renewable Energy (K.A.CARE).
	
	\bibliographystyle{IEEEtran}
	\bibliography{References}

\begin{thebibliography}{10}
\providecommand{\url}[1]{#1}
\csname url@samestyle\endcsname
\providecommand{\newblock}{\relax}
\providecommand{\bibinfo}[2]{#2}
\providecommand{\BIBentrySTDinterwordspacing}{\spaceskip=0pt\relax}
\providecommand{\BIBentryALTinterwordstretchfactor}{4}
\providecommand{\BIBentryALTinterwordspacing}{\spaceskip=\fontdimen2\font plus
\BIBentryALTinterwordstretchfactor\fontdimen3\font minus
  \fontdimen4\font\relax}
\providecommand{\BIBforeignlanguage}[2]{{%
\expandafter\ifx\csname l@#1\endcsname\relax
\typeout{** WARNING: IEEEtran.bst: No hyphenation pattern has been}%
\typeout{** loaded for the language `#1'. Using the pattern for}%
\typeout{** the default language instead.}%
\else
\language=\csname l@#1\endcsname
\fi
#2}}
\providecommand{\BIBdecl}{\relax}
\BIBdecl

\bibitem{he2018lte}
L.~He, Z.~Yan, and M.~Atiquzzaman, ``Lte/lte-a network security data collection
  and analysis for security measurement: A survey,'' \emph{IEEE Access},
  vol.~6, pp. 4220--4242, 2018.

\bibitem{giluka2018enhanced}
S.-I. Sou, M.-R. Li, S.-H. Wang, and M.-H. Tsai, ``File distribution via
  proximity group communications in lte-advanced public safety networks,''
  \emph{Computer Networks}, vol. 149, pp. 93--101, 2019.

\bibitem{xu2018hybrid}
Y.-C. Wang and C.-C. Huang, ``Efficient management of interference and power by
  jointly configuring abs and drx in lte-a hetnets,'' \emph{Computer Networks},
  vol. 150, pp. 15--27, 2019.

\bibitem{ref2}
{N}okia~{S}iemens {N}etwork, ``Signaling is growing 50\% faster than data
  traffic,'' 2012.

\bibitem{razavi2014reducing}
S.~M. Razavi and D.~Yuan, ``Reducing signaling overhead by overlapping tracking
  area list in lte,'' in \emph{Wireless and Mobile Networking Conference
  (WMNC), 2014 7th IFIP}.\hskip 1em plus 0.5em minus 0.4em\relax IEEE, 2014,
  pp. 1--7.

\bibitem{aqeeli2017dynamic}
E.~Aqeeli, A.~Moubayed, and A.~Shami, ``Dynamic son-enabled location management
  in lte networks,'' \emph{IEEE Transactions on Mobile Computing}, 2017.

\bibitem{bonami2008algorithmic}
P.~Bonami, L.~T. Biegler, A.~R. Conn, G.~Cornu{\'e}jols, I.~E. Grossmann, C.~D.
  Laird, J.~Lee, A.~Lodi, F.~Margot, N.~Sawaya \emph{et~al.}, ``An algorithmic
  framework for convex mixed integer nonlinear programs,'' \emph{Discrete
  Optimization}, vol.~5, no.~2, pp. 186--204, 2008.

\bibitem{abido2009multiobjective}
M.~Abido, ``Multiobjective particle swarm optimization for
  environmental/economic dispatch problem,'' \emph{Electric Power Systems
  Research}, vol.~79, no.~7, pp. 1105--1113, 2009.

\bibitem{wong2000location}
V.-S. Wong and V.~C. Leung, ``Location management for next-generation personal
  communications networks,'' \emph{IEEE network}, vol.~14, no.~5, pp. 18--24,
  2000.

\bibitem{munir2018secure}
K.~Munir, E.~Zahoor, R.~Rahim, X.~Lagrange, and J.-H. Lee, ``Secure and
  fault-tolerant distributed location management for intelligent 5g wireless
  networks,'' \emph{IEEE Access}, 2018.

\bibitem{ref14}
D.~Gu and S.~Rappaport~Stephen, ``Mobile user registration in cellular systems
  with overlapping location areas,'' in \emph{1999 IEEE 49th Vehicular
  Technology Conference (VTC'99)}, vol.~1, Jul 1999, pp. 802--806 vol.1.

\bibitem{ref13}
S.-R. Yang, Y.-C. Lin, and Y.-B. Lin, ``Performance of mobile
  telecommunications network with overlapping location area configuration,''
  \emph{IEEE Transactions on Vehicular Technology}, vol.~57, no.~2, pp.
  1285--1292, March 2008.

\bibitem{ref15}
K.-H. Chiang and N.~Shenoy, ``A 2-d random-walk mobility model for
  location-management studies in wireless networks,'' \emph{IEEE Transactions
  on Vehicular Technology}, vol.~53, no.~2, pp. 413--424, March 2004.

\bibitem{widjaja2009comparison}
I.~Widjaja, P.~Bosch, and H.~La~Roche, ``Comparison of mme signaling loads for
  long-term-evolution architectures,'' in \emph{Vehicular Technology Conference
  Fall (VTC 2009-Fall), 2009 IEEE 70th}.\hskip 1em plus 0.5em minus 0.4em\relax
  IEEE, 2009, pp. 1--5.

\bibitem{kunz2010minimizing}
A.~Kunz, T.~Taleb, and S.~Schmid, ``On minimizing serving gw/mme relocations in
  lte,'' in \emph{Proceedings of the 6th International Wireless Communications
  and Mobile Computing Conference}.\hskip 1em plus 0.5em minus 0.4em\relax ACM,
  2010, pp. 960--965.

\bibitem{taleb2014supporting}
T.~Taleb, K.~Samdanis, and A.~Ksentini, ``Supporting highly mobile users in
  cost-effective decentralized mobile operator networks,'' \emph{IEEE
  Transactions on Vehicular Technology}, vol.~63, no.~7, pp. 3381--3396, 2014.

\bibitem{ref16}
S.~Razavi, D.~Yuan, F.~Gunnarsson, and J.~Moe, ``Exploiting tracking area list
  for improving signaling overhead in {LTE},'' in \emph{2010 IEEE 71st
  Vehicular Technology Conference (VTC'10-Spring)}, May 2010, pp. 1--5.

\bibitem{ref17}
S.~Razavi, D.~Yuan, F.~Gunnarsson, and J.~Moe, ``Dynamic tracking area list
  configuration and performance evaluation in {LTE},'' in \emph{2010 IEEE
  GLOBECOM Workshops (GC'10 Wkshps)}, Dec 2010, pp. 49--53.

\bibitem{ref18}
S.~Modarres~Razavi and D.~Yuan, ``Mitigating mobility signaling congestion in
  {LTE} by overlapping tracking area lists,'' in \emph{Proceedings of the 14th
  ACM International Conference on Modeling, Analysis and Simulation of Wireless
  and Mobile Systems}, ser. MSWiM '11.\hskip 1em plus 0.5em minus 0.4em\relax
  New York, NY, USA: ACM, 2011, pp. 285--292.

\bibitem{eberhart_new_1995}
R.~C. Eberhart and J.~Kennedy, ``A new optimizer using particle swarm theory,''
  in \emph{Proceedings of the sixth international symposium on micro machine
  and human science}, vol.~1.\hskip 1em plus 0.5em minus 0.4em\relax New York,
  {NY}, 1995, pp. 39--43.

\bibitem{du2015using}
J.~Du, L.~Zhao, J.~Xin, J.-M. Wu, and J.~Zeng, ``Using joint particle swarm
  optimization and genetic algorithm for resource allocation in td-lte
  systems,'' in \emph{Heterogeneous Networking for Quality, Reliability,
  Security and Robustness (QSHINE), 2015 11th International Conference
  on}.\hskip 1em plus 0.5em minus 0.4em\relax IEEE, 2015, pp. 171--176.

\bibitem{wu2006energy}
X.~Wu, S.~Lei, W.~Jin, J.~Cho, and S.~Lee, ``Energy-efficient deployment of
  mobile sensor networks by pso,'' in \emph{Asia-Pacific Web Conference}.\hskip
  1em plus 0.5em minus 0.4em\relax Springer, 2006, pp. 373--382.

\bibitem{ab2009wireless}
N.~A.~B. Ab~Aziz, A.~W. Mohemmed, and M.~Y. Alias, ``A wireless sensor network
  coverage optimization algorithm based on particle swarm optimization and
  voronoi diagram,'' in \emph{Networking, Sensing and Control, 2009. ICNSC'09.
  International Conference on}.\hskip 1em plus 0.5em minus 0.4em\relax IEEE,
  2009, pp. 602--607.

\bibitem{hashim2015_L1_SISO}
H.~A. Hashim, S.~El-Ferik, and M.~A. Abido, ``A fuzzy logic feedback filter
  design tuned with pso for l1 adaptive controller,'' \emph{Expert Systems with
  Applications}, vol.~42, no.~23, pp. 9077--9085, 2015.

\bibitem{abido_optimal_2002}
M.~A. Abido, ``Optimal power flow using particle swarm optimization,''
  \emph{International Journal of Electrical Power \& Energy Systems}, vol.~24,
  no.~7, pp. 563--571, 2002.

\bibitem{abido2010multiobjective}
M.~A. Abido, ``Multiobjective particle swarm optimization with nondominated
  local and global sets,'' \emph{Natural Computing}, vol.~9, no.~3, pp.
  747--766, 2010.

\bibitem{sun2018attack}
M.~Iqbal, M.~Naeem, A.~Anpalagan, N.~N. Qadri, and M.~Imran, ``Multi-objective
  optimization in sensor networks: Optimization classification, applications
  and solution approaches,'' \emph{Computer Networks}, vol.~99, pp. 134--161,
  2016.

\bibitem{hashim2017_L1_MIMO}
H.~A. Hashim, S.~El-Ferik, B.~O. Ayinde, and M.~A. Abido, ``Optimal tuning of
  fuzzy feedback filter for l1 adaptive controller using multi-objective
  particle swarm optimization for uncertain nonlinear mimo systems,''
  \emph{arXiv preprint arXiv:1710.05423}, 2017.

\bibitem{eltoukhy1}
A.~E. Eltoukhy, Z.~Wang, F.~T. Chan, and S.~Chung, ``Joint optimization using a
  leader--follower stackelberg game for coordinated configuration of stochastic
  operational aircraft maintenance routing and maintenance staffing,''
  \emph{Computers \& Industrial Engineering}, vol. 125, pp. 46--68, 2018.

\bibitem{eltoukhy3}
A.~E. Eltoukhy, F.~T. Chan, S.~H. Chung, B.~Niu, and X.~Wang, ``Heuristic
  approaches for operational aircraft maintenance routing problem with maximum
  flying hours and man-power availability considerations,'' \emph{Industrial
  Management \& Data Systems}, vol. 117, no.~10, pp. 2142--2170, 2017.

\bibitem{eltoukhy5}
A.~E. Eltoukhy, Z.~Wang, F.~T. Chan, and X.~Fu, ``Data analytics in managing
  aircraft routing and maintenance staffing with price competition by a
  stackelberg-nash game model,'' \emph{Transportation Research Part E:
  Logistics and Transportation Review}, vol. 112, pp. 143--168, 2019.

\bibitem{kamel2017online}
M.~Kamel, S.~Elkatatny, M.~Mysorewala, M.~Elshafei, and A.~Ai-Majed, ``Online
  control of stick-slip in rotary steerable drilling,'' in \emph{2017 9th
  IEEE-GCC Conference and Exhibition (GCCCE)}.\hskip 1em plus 0.5em minus
  0.4em\relax IEEE, 2017, pp. 1--7.

\bibitem{kamel2017quad}
M.~Kamel, M.~Abido, and M.~Elshafei, ``Quad-rotor directional steering system
  controller design using gravitational search optimization,''
  \emph{Intelligent Automation \& Soft Computing}, pp. 1--11, 2017.

\bibitem{Hash_2015_Comparative}
H.~A. Hashim and M.~Abido, ``Fuzzy controller design using evolutionary
  techniques for twin rotor mimo system: a comparative study,''
  \emph{Computational intelligence and neuroscience}, vol. 2015, no.~9, p.~11,
  2015.

\bibitem{mohamed2014improved}
H.~A.~H. Mohamed, ``Improved robust adaptive control of high-order nonlinear
  systems with guaranteed performance,'' \emph{M. Sc, King Fahd University Of
  Petroleum \& Minerals}, vol.~1, 2014.

\bibitem{Hashim_2016_Network}
H.~A. Hashim, B.~Ayinde, and M.~Abido, ``Optimal placement of relay nodes in
  wireless sensor network using artificial bee colony algorithm,''
  \emph{Journal of Network and Computer Applications}, vol.~64, pp. 239--248,
  2016.

\bibitem{zitzler1999evolutionary}
E.~Zitzler, \emph{Evolutionary algorithms for multiobjective optimization:
  Methods and applications}.\hskip 1em plus 0.5em minus 0.4em\relax Citeseer,
  1999, vol.~63.

\end{thebibliography}

	\vspace{100pt}

	\section*{AUTHOR INFORMATION}
    \vspace{10pt}
	{\bf Hashim A. Hashim} is a Ph.D. candidate and a Teaching and Research Assistant in Robotics and Control, Department of Electrical and Computer Engineering at the University of Western Ontario, ON, Canada.\\
	His current research interests include stochastic and deterministic filters on SO(3) and SE(3), control of multi-agent systems, control applications and optimization techniques.\\
	\underline{Contact Information}: \href{mailto:hmoham33@uwo.ca}{hmoham33@uwo.ca}.
	\vspace{30pt}\\

	{\bf Mohammad A. Abido}  received the B.Sc. (Honors with first class) and M.Sc. degrees in EE from Menoufia University, Shebin El-Kom, Egypt, in 1985 and 1989, respectively, and the Ph.D. degree from King Fahd University of Petroleum and Minerals (KFUPM), Dhahran, Saudi Arabia, in 1997. He is currently a \textbf{Distinguished University Professor} at KFUPM.\\
	His research interests are power system control and operation and renewable energy resources integration to power systems. Dr. Abido is the recipient of KFUPM\textbf{ Excellence in Research Award}, 2002, 2007 and 2012, KFUPM \textbf{Best Project Award}, 2007 and 2010, \textbf{First Prize Paper Award} of the Industrial Automation and Control Committee of the IEEE Industry Applications Society, 2003, \textbf{Abdel-Hamid Shoman Prize for Young Arab Researchers} in Engineering Sciences, 2005, \textbf{Best Applied Research Award} of 15thGCC-CIGRE Conference, Abu-Dhabi, UAE, 2006, and \textbf{Best Poster Award}, International Conference on Renewable Energies and Power Quality (ICREPQ’13), Bilbao, Spain, 2013. Dr. Abido has been awarded \textbf{Almarai Prize for Scientific Innovation 2017-2018, Distinguished Scientist, Saudi Arabia, 2018 and Khalifa Award for Education 2017-2018, Higher Education, Distinguished University Professor in Scientific Research}, Abu Dhabi, UAE, 2018. Dr. Abido has published two books and more than 350 papers in reputable journals and international conferences. He participated in 60+ funded projects and supervised 50+ MS and PhD students.
	\vspace{167pt}

\end{document}